\documentclass[a4paper,11pt]{article}
\pdfoutput=1 

\usepackage{jcappub} 

\usepackage[T1]{fontenc} 
\usepackage{journals}
\usepackage{bm}
\usepackage{multirow}

\usepackage[english]{babel}
\usepackage[utf8x]{inputenc}
\usepackage[T1]{fontenc}

\usepackage[a4paper,top=3cm,bottom=2cm,left=3cm,right=3cm,marginparwidth=1.75cm]{geometry}

\usepackage{amsmath}
\usepackage{graphicx}
\usepackage{amssymb}
\usepackage[colorinlistoftodos]{todonotes}

\def \mpc {\mbox{\rm Mpc}}

\title{Calibration of systematics in constraining modified gravity models with galaxy cluster mass profiles}
\author[1]{L. Pizzuti,}
\author[2,3]{B. Sartoris,}
\author[2,3,4,5]{S. Borgani,}
\author[2,3]{A. Biviano.}

\affiliation[1]{OAVdA - Osservatorio Astronomico della Regione Autonoma Valle d'Aosta, Loc. Lignan 39, I-11020, Nus, Italy}
\affiliation[2]{INAF - Osservatorio Astronomico di Trieste,\\ Via Tiepolo 11, I-34143 Trieste, Italy}
\affiliation[3]{IFPU - Institute for Fundamental Physics of the Universe, Via Beirut 2, 34014 Trieste, Italy}
\affiliation[4]{ INFN - Sezione di Trieste,\\ Via Valerio 2, I-34127 Trieste, Italy}
\affiliation[5]{Dipartimento di Fisica, Sezione di Astronomia, Universit\`a di Trieste,\\ Via Tiepolo 11, I-34143 Trieste, Italy}

\emailAdd{pizzuti@oavda.it}

\abstract{Joint lensing and dynamical mass profile determinations of galaxy clusters are an excellent tool to constrain modification of gravity at cosmological scales. However, search for tiny departures from General Relativity (GR) calls for an accurate control of the systematics affecting the method. In this analysis we concentrate on the systematics in the reconstruction of mass profiles from the dynamics of cluster member galaxies, while assuming that lensing provides unbiased mass profile reconstructions. In particular, in the case study of linear $f(R)$ gravity, we aim at verifying whether in realistic simulations of cluster formation a spurious detection of departure from GR can be detected due to violation of the main assumptions (e.g. dynamical equilibrium and spherical symmetry) on which the method is based. We aim at identifying and calibrating the impact of those systematics by analyzing a set of Dark Matter halos taken from $\Lambda$CDM N-body cosmological simulations performed with the GADGET-3 code.
 We evaluate how the constraints on the additional degree of freedom $m_{f_{R}}$ suffer the lack of dynamical relaxation and departures from spherical symmetry. 
 If no selection criteria are applied, $\sim 60\%$ of clusters in a $\Lambda$CDM Universe (where GR is assumed) produce a spurious detection of modified gravity. We find that the probability of finding cluster in agreement with GR predictions $P_{GR}$ mainly depends on the properties of the halo's projected phase-space and on shape orientation of the cluster along the line-of-sight projection. We further define two observational criteria which correlate with $P_{GR}$ and which can be used to select, among a generic population of galaxy clusters in the local Universe, those objects that are more suitable for the application of the proposed method. In particular, we find that according to these criteria the percentage of spurious detection can be lowered down to $\sim 20\%$ in the best case.  We discuss how $P_{GR}$ changes when a simplified treatment of the chameleon screening mechanism is considered. Our results are relevant in view of the availability of precise measurements of lensing mass profiles from imaging data and dynamics mass profiles from spectroscopic data  that will be available with the next generation surveys.
}

\begin{document}
\maketitle
\flushbottom
\section{Introduction}
\label{sec:introd}
In the cosmological concordance model, or $\Lambda$CDM model, the late-time accelerated expansion of the universe (refs. \cite{Reiss01,perlmutter99}) is explained by introducing an additional constant term in the Einstein equations of General Relativity (GR hereafter). Despite this cosmological constant $\Lambda$ cannot be naturally described in the framework of standard physics, the concordance model remains consistent with observations, even with the large improvement in the quantity and quality of available data (e.g. ref. \cite{Planck18r}) .\\
Among the several viable alternatives proposed to investigate the origin of the acceleration, an interesting possibility is to modify the theory of General Relativity (GR hereafter), assumed in the Standard Cosmological model. Modification of gravity on large scales can be modeled to mimic the effect
of a cosmological constant in the background expansion; however departures from GR introduce new degrees of freedom which can substantially change the evolution of cosmological structures from the prediction of  $\Lambda$CDM model (see e.g. ref. \cite{Ishak} and references therein for a review). This allows for the possibility to constrain possible modification of gravity with several cosmological probes, including Cosmic Microwave Background anisotropies (e.g. ref. \cite{Planckmod}), gravitational waves (e.g. refs. \cite{Abbott17,Belgacem19}) and large scale structure probes (e.g. refs. \cite{Davies19,Ya2006,J2012,Mancini19}).
Among the broad range of gravity tests at cosmological scales, galaxy clusters offer an interesting field of investigation (e.g. refs. \cite{Cataneo16,Mack11,Sakstein16}); in particular, cluster mass profile reconstructions using complementary relativistic (i.e. photons) and non-relativistic (i.e. diffuse X-ray emitting gas and member galaxies) tracers represent an excellent tool to put competitive constraints on modified gravity models (see e.g. refs. \cite{Terukina14,Pizzuti16}).

In a galaxy cluster, under the assumption of dynamical relaxation, the motion of galaxies within the cluster is governed by the Jeans' equation, which relates the velocity dispersion and the number density profiles of the galaxies to the underlying gravitational potential. Since galaxies are non relativistic objects (typical velocity dispersion is of the order of $\sim 10^3\,km/s \ll c$, see e.g. \cite{Biviano01}) they perceive only the time-time component of the metric, expressed by the potential $\Phi$. On the other hand, photons, due to the conformal structure of Maxwell equations, are sensitive to both $\Phi$ and $\Psi$. We thus have two distinct probes of the same physical quantity (the mass profile) which are differently related to the metric potentials. A combined analysis of lensing and dynamics information can be used to constraints deviation from GR with galaxy clusters.

A popular class of viable modified gravity models is the $f(R)$ theory of gravity, where a function of the Ricci scalar $R$ is added to the Einstein-Hilbert action of General Relativity. An extra degree of freedom $f_{R}=df/dR$, called the scalaron, mediates a Yukawa-type fifth force which enhances gravity over scales smaller than the characteristic interaction range $\lambda=1/m_{f_{R}}$.
The larger is the mass of the scalaron $m_{f_{R}}$ the smaller is the departure from standard gravity.

Ref. \cite{Pizzuti17} performed a joint lensing and dynamics analysis on the data of two relaxed galaxy clusters MACS J1206.2-0847
(hereafter MACS 1206) at redshift $z=0.44$, and RXC J2248.7-4431
(hereafter RXJ 2248) at $z=0.35$, aimed at constraining the interaction range $\lambda$ in the framework of linear $f(R)$ gravity. An interesting behavior has been found in the case of the cluster RXJ 2248, where a slight preference of  MG with respect to GR suggests that some effects, whose contribution can be negligible in the framework of General Relativity, may become relevant source of systematics when searching for small deviations, such as those induced by the additional degrees of freedom in non-standard gravity. 
The result calls for a solid control of the assumptions on which the method relies in order to draw robust conclusions. In particular, deviations from spherical symmetry and lack of dynamical relaxation state of the cluster introduce systematic effects which can produce spurious detection of modified gravity. While departures from spherical symmetry affect both lensing and dynamics mass profile determinations, relaxation is a feature related to the dynamics analysis only. The calibration of these systematics has important implications in view of upcoming and future large imaging and spectroscopic surveys (e.g Euclid, LSST), that will deliver lensing and kinematic mass reconstruction for a large number of galaxy clusters.

In order to estimate the impact of the above mentioned effects in the dynamics mass profile determination, in this paper we analyze a set of 58 $\Lambda$CDM Dark Matter (DM hereafter) cluster-size halos taken at $z=0$ from zoomed-in re-simulations of 29 Lagrangian regions, extracted from a parent cosmological DM-only simulation (see e.g. refs. \cite{Biffi16,Planellas13} and references therein). The parent simulation is performed with the parallel code GADGET-3 \cite{Springel05}.


We reconstruct the dynamics mass profiles of the simulated clusters in the framework of linear $f(R)$ gravity by using a modified version of the \emph{MAMPOSSt} method of ref. \cite{Mamon01} (\emph{MG-MAMPOSSt}), aiming at constraining the scalaron mass  $m_{f_R}$. It is worth to notice that in a realistic scenario $f(R)$ models exhibit an environment-dependent $m_{f_R}$ which suppresses GR modification at small scales in high density regions (chameleon screening, see e.g. ref. \cite{Brax08}). The simple linearized framework of modified gravity has to be regarded as a test case where the additional (constant) free parameter $m_{f_{R}}$ is used to quantify the magnitude of the deviations from the assumption of GR.  We will discuss possible effects due to a non-linear chameleon screening in Section \ref{sec:screen}.

 Assuming a Navarro-Frenk-White (NFW hereafter) model for the matter density perturbations, which has been shown to provide an good description for simulated DM halos as well for real cluster's density profiles ( e.g. refs. \cite{Peirani17,Umetsu16}), we quantify spurious departures from the GR expectation as the contribution induced by systematics; we further introduce two observational criteria related to deviation from the main assumptions of the method and which correlate with the probability of finding cluster which do not exclude GR. These criteria are based on the projected distribution of velocities and position of cluster member galaxies (projected phase space), which can be directly extracted from observational data.

The paper is organized as follows: in Section \ref{sec:theo} we review the basic notions of $f(R)$ models, in Section \ref{sec:mam} we briefly discuss dynamical mass profile determinations through the Jeans' equation and the \emph{MG-MAMPOSSt} method. Section \ref{sec:sim} is devoted to describe the simulated sample of halos used in the analysis; in Section \ref{sec:results} we present our results, which are further discussed in Section \ref{sec:conc}, where we draw our main conclusions.
\section{$f(R)$ theories of gravity}

\label{sec:theo}
$f(R)$ gravity, initially proposed by ref. \cite{Buch01}, is one of the most widely investigated modified gravity model. Despite their relative simplicity, $f(R)$ theories encapsulate a broad range of features which satisfy local test of gravity and avoid theoretical instabilities (see e.g. refs. \cite{Song2007,Hu2007,Pogosian10}); in addition, for large values of the modification $f(R)$ models are capable of explaining the late time accelerated expansion of the Universe\footnote{Note that this is not true for small departures from GR, where pure scalar-tensor theories cannot produce the observed cosmic acceleration without the additional contribution of other dark energy sources (see e.g. ref. \cite{Lombriser17})}. As such, they provide an excellent framework within which studying alternatives to GR in a cosmological context.

In the Jordan frame, the total action of a generic $f(R)$ model is given by: 
\begin{equation}
S\,=\,\frac{1}{16\pi G}\int{\sqrt{-g}[R+f(R)]d^{4}x}+S_{m}[\Psi_{m},g_{\mu\nu}],\label{eq:fr}
\end{equation}
where  $f(R)$ is a generic non-linear function of the Ricci scalar which reduces to $\Lambda$CDM for $f(R)=\mathrm{const}\equiv-2\Lambda$. $S_{m}$ is the action of the matter field $\Psi_{m}$.

The additional degree of freedom of the theory can be expressed in terms of:
\[
f_{R}=\frac{df(R)}{dR},
\]
dubbed as the scalaron field, with a characteristic mass given by (e.g. ref. \cite{Hu2007}):
\begin{equation}
m_{f_{R}}^{2}=\frac{1}{3}\left(\frac{1+f_{R}}{f_{RR}}-R\right),\label{mass}
\end{equation}
Note that the inverse of the scalaron mass $\lambda=1/m_{f_{R}}$ describes the typical interaction range of the additional Yukawa-type force. $f_{,RR}=df_{R}/dR \ge 0$ is the second derivative of $f(R)$.

In the quasistatic approximation, which is valid for scales well inside the horizion, and assuming that fluctuations in the curvature remain everywhere small (i.e. linear $f(R)$ gravity), we can derive an expression for the modified time-time gravitational potential $\Phi$. Following e.g. ref. \cite{Pizzuti17}, one obtains:
 \begin{equation}
     \Phi(r)=(1+f_R)^{-1}\left[G\int_{r_{0}}^{r}{\frac{dx}{x^{2}}M(x)} +\frac{1}{3}\phi_{mg}(r,m_{f_R})\right],\label{potfr}
 \end{equation}
where $\phi_{mg}(r,m_{f_R})$ expresses departures from GR in terms of the scalaron mass and $M(r)$ is the mass enclosed within a radius $r$. Working in linear $f(R)$ gravity is equivalent to say that $m_{f_R}$ is constant over the scale analysed.

From the geodesics equation it can be seen that the motion of non relativistic objects, as galaxies in clusters, is determined by the time-time gravitational potential $\Phi$;  we can define an effective dynamical mass as the mass sourced by $\Phi$ according to:
\begin{equation}\label{eq:mdyn}
\frac{d\Phi}{dr}=\frac{G\,M_{dyn}(r)}{r^2},
\end{equation} where $M_{dyn}=M(r)+(r^2/3)\left[d\phi_{mg}(r,m_{f_R})/dr\right]$ is the sum of the mass measured in standard gravity and the contribution of the fifth force.

As for the matter density perturbations $\delta \rho$ we adopt the NFW parametrization of ref. \cite{Navarro}:
\begin{equation}\label{eq:NFWdens}
    \delta\rho=\frac{\rho_0}{r/r_s(1+r/rs)^2},
\end{equation}
a double-power law radial profile, proportional to $r^{-1}$ in the innermost region and to $r^{-3}$ at large radii. $\rho_0,\, r_s$ are two parameters defining the shape of the profile. In particular $\rho_0$ is the characteristic density of the halo, while $r_s\equiv r_{-2}$ is the scale radius at which the logarithmic derivative of the profile equals $-2$. Ref. \cite{Navarro} shows that eq. \eqref{eq:NFWdens}  provides a good fit for Dark Matter halos in N-body simulations over two decades in radius.\\
Given its relative simplicity, the NFW profile has been widely adopted in literature to
accurately describe the total mass profile of gravitationally bound structures such as galaxy clusters, both in the analysis of cosmological simulations and in the reconstruction of mass profiles from observations (e.g. refs. \cite{Biviano01,Umetsu16}). However, it is worth to notice that some discrepancies between different sets of simulations and
observations have been found regarding the inner shape of Dark matter halo mass profiles.
In particular, the analysis of ref. \cite{navarro04} on high resolution cosmological N-body simulations showed that
the central region of CDM halos follows a profile better described by an exponential law (Einasto profile); more recent studies (see e.g. refs. \cite{Pontzen12}) indicate that the presence of baryons can either flatten or steepen the Dark Matter halo profile with respect to a pure CDM paradigm, depending on the details of the physical processes (cooling, star formation and energy feedback processes) related to the presence of baryons. Nevertheless, the mass distribution in the innermost region of galaxy clusters cannot be determined with the method discussed in this paper, as the dynamics in the center is no more described by the Jeans' equation. Thus, as we will see below, the central mass profile of the halos is excluded in our analysis.\\
Moreover, despite the controversies about the inner shape are still open, the NFW profile remains a good model to characterize
the overall distribution of matter (baryonic and dark) in cosmological structures.

With this assumption, in eq. \eqref{potfr} we have:
\begin{equation*}
M(r)\equiv M_{NFW}(r)=
\end{equation*}
\begin{equation} \label{eq:NFWmass}
=M_{200}\left[\log(1+r/r_{s})-r/r_{s}(1+r/r_{s})^{-1}\right]\times\left[\log(1+c_{200})-c_{200}/(1+c_{200})\right]^{-1},
\end{equation}
and

\begin{equation*}
\phi_{mg}(r)= 3\frac{2\pi G\rho_0}{r}r_s^3\left\{e^{-m_{f_R}(r_s+r)}\left[{\rm Ei}(m_{f_R}\,r_s)-{\rm Ei}(m_{f_R}(r_s+r))\right]\right.
\end{equation*}
\begin{equation}\label{eq:modmass}
\left.-e^{m_{f_R}(r_s+r)}{\rm Ei}\left[-m_{f_R}(r_s+r)\right]+e^{m_{f_R}(r_s-r)}{\rm Ei}(-m_{f_R}\,r_s)\right\},
\end{equation}
where ${\rm Ei}(x)$ is the exponential integral function.
Instead of $\rho_0$, in eq. \eqref{eq:NFWmass} we have used the parameter $r_{200}$, defined as the radius at which the density is 200 times the critical density of the Universe. $c_{200}=r_{200}/r_s$ is called concentration, $M_{200}=(4/3)\pi200\delta_c r_{200}^3$ is the total mass enclosed within $r_{200}$.

In $f(R)$ gravity, due to the conformal structure of the model, photon propagation is not affected by the fifth force contribution except of a conformal rescaling of the gravitational constant $G_{eff}=G/(1+f_{R})$. Thus, lensing mass profile reconstructions are sensitive to $M_{NFW}$ only plus second-order corrections (see e.g. refs. \cite{Schmidt10,Pizzuti17}), and can be used as an additional information on the GR parameters $r_s,r_{200}$. In this work we will not consider explicitly systematics in lensing analyses. As a consistency check, in Appendix \ref{app:const} we derive the expected constraints on $m_{f_R}$ obtainable for a synthetic cluster for which all the assumptions are met, generated with the method of ref. \cite{Pizzuti19}, combining the likelihood of the dynamical mass  profile reconstruction (see Section \ref{sec:mam}) with simulated probability distributions $P_L(r_s,r_{200})$ expected from a lensing mass reconstruction. 

\section{Dynamical mass profiles with MG-\emph{MAMPOSSt}}
\label{sec:mam}
Under the assumption that galaxies in clusters are collisionless tracers of the total time-time gravitational potential $\Phi$ (i.e. dynamical relaxation), the dynamics is governed by the Jeans' equation (Jeans, 1919) which, for a spherical symmetric system is given by:
\begin{equation}
\frac{d\left[\nu(r)\sigma^2_{r}(r)\right]}{dr}+2\beta(r)\frac{\nu(r)\sigma_{r}^{2}(r)}{r}=-\nu(r)\frac{d\Phi(r)}{dr}\label{eq:Jeans},
\end{equation}
in the above expression, $\sigma^2_{r}$ is the velocity dispersions corresponding to the the radial component $v_r$, $\nu(r)$ is the number density profile of the tracers. We have defined the velocity anisotropy profile $\beta(r)$ as the ratio:
\begin{equation}\label{eq:betadef}
\beta(r)=1-\frac{\sigma^2_{\theta}+\sigma^2_{\phi}}{2\sigma^2_r},
\end{equation}
where $\sigma^2_{\theta,\phi}$ are the velocity dispersions for the components $\,v_\theta,\,v_\phi$ respectively; since in spherical symmetry $\sigma^2_{\theta}=\sigma^2_{\phi}$,  the velocity anisotropy reduces to:
\begin{equation}
\beta(r)=1-\frac{\sigma^2_\theta}{\sigma^2_r}.
\end{equation}

Thus, measurements of the number density profile $\nu$, of the velocity dispersion $\sigma_r$ and of the velocity anisotropy $\beta(r)$ allow us to constrain the gravitational potential through eq. \eqref{eq:Jeans} which is further related to the effective dynamical mass (see eq. \eqref{eq:mdyn}).

It is important to point out that the velocity anisotropy profile is in general unknown and should be either modeled with parametric profiles based on study of cosmological simulations (see e.g. refs. \cite{Mamon01,Hansen2006,mamon10}) or reconstructed via non-parametric approaches based on the inversion of the Jeans' equation (e.g. ref. \cite{Host2}). The ignorance over $\beta(r)$ represents a source of systematics in mass profile determination through the Jeans' analysis.

In this paper we implement a parametric reconstruction of the velocity anisotropy profile. The chosen models of $\beta(r)$ are fitted along with the total gravitational potential by using the \emph{MAMPOSSt} method of ref. \cite{Mamon01}.\\
\emph{MAMPOSSt} stands for \emph{Modelling Anisotropy and Mass Profiles of Observed Spherical Systems} and it is a technique to infer the total cluster mass profile by solving the Jeans' equation. The advantage of \emph{MAMPOSSt} relies on the fact that it requires only projected information of galaxy velocities and positions. More in details, the method performs a Maximum Likelihood fit in the projected phase space of cluster galaxies, defined as the plane $(R,v_{los})$, where R is the projected position of a galaxy with respect to the cluster center and $v_{los}$ is the velocity measured along the line of sight in the rest frame of the cluster. \emph{MAMPOSSt} works under the assumption of a 3-dimensional Gaussian distribution for the velocities of the tracers. It is worth to point out that the method can be generally applied given any model of the 3D velocity distribution; the choice of a Gaussian is the simplest, but the code has been extensively tested and verified to work quite well on halos drawn from cosmological simulations in which the velocity distribution along the l.o.s. can have significant deviations from Gaussianity (see ref. \cite{Mamon01}).
In our version of the code we assume a parametric form of the gravitational potential, of the number density profile of the tracers and of the velocity anisotropy to obtain the values of the free parameters that better fit the data. Note that the number density profile $\nu(r)$ can in general be measured directly from the phase space. As such, we perform a Maximum Likelihood fit which does not require the binning of data (see e.g. ref. \cite{Biviano01}) over the distribution of tracers in our sample of phase spaces; we assume a projected NFW model, obtained by integrating eq. \eqref{eq:NFWdens} along the l.o.s.,  where the only free parameter is the scale radius $r_{\nu}$ of the profile. Note that we have changed subscript index from the scale radius of the mass profile $r_s$ since in general the distribution of galaxies in clusters is different from the distribution of the total matter density (see e.g. refs. \cite{Biviano06,budzynski12}) The best fit value of $r_{\nu}$ is used as input in the \emph{MAMPOSSt} procedure.\\
Ref. \cite{Pizzuti17} developed an extension of \emph{MAMPOSSt} (\emph{MG-MAMPOSSt} hereafter) by introducing a generic parametric form for the gravitational potential $\Phi$. In the case of a NFW density profile for the matter perturbation the expression of $\Phi$ reads: 
\begin{equation}
\label{eq:phimod}
\Phi(r)=h_{1}\left[G\int_{r_{0}}^{r}{\frac{dx}{x^{2}}M_{NFW}(x,r_s,r_{200})} +2Q^2\phi_{mg}(x,r_s,r_{200},m,S)\right],
\end{equation}
which accounts for a broad range of viable modified gravity models, including the $f(R)$ case. The parameter array now contains 4 additional parameters,
which are the mass of the scalaron $m\equiv m_{f_R}$\footnote{from now on we drop the subscript ${f_R}$ in the scalaron mass}, the conformal rescaling $h_{1}$, the coupling factor  $Q$  and the screening
radius $S$, under the assumption of an instantaneous transition between the screened and non-screened regime. Comparing eq. \eqref{eq:phimod} and eq. \eqref{potfr} of Section \ref{sec:theo}, it is straightforward to see that in linear $f(R)$ gravity $2Q^{2}=1/3$, $h_{1}=1/(1+f_{,R})\simeq 1$, $S\sim 0$. Thus, the free parameters in the gravitational potential are the scale radius $r_s$, the virial radius $r_{200}$ and the mass of the scalaron field $m$. As for the velocity anisotropy, we consider 4 models implemented in the \emph{MG-MAMPOSSt} code: \\
the \textbf{constant anisotropy model $\text{''C''}$} 
\begin{equation}\label{eq:betac}
\beta(r)=\beta_C,
\end{equation}
the \textbf{Mamon\&Lokas model $\text{''ML''}$} of ref. \cite{MamLok05}, which has been shown to provide a good fit to the average cluster-size halos anisotropy profile over a set of cosmological simulations (see e.g ref. \cite{mamon10})
\begin{equation}\label{eq:betaml}
\beta(r)_{ML}=\frac{1}{2}\frac{r}{r+r_{\beta}},
\end{equation}
where $r_{\beta}$ is a characteristic scale radius; \\
the \textbf{Tiret model  $\text{''T''}$} of ref. \cite{Tiret2007}
\begin{equation}\label{eq:betat}
\beta(r)_T=\beta_{\infty}\frac{r}{r+r_{c}},
\end{equation}
a generalized version of the  $\text{''ML''}$ which tends to $\beta_{\infty}$ at large radii;\\
the \textbf{Opposite model  $\text{''O''}$}
\begin{equation}\label{eq:betao}
\beta(r)_O=\beta_{\infty}\frac{r-r_{c}}{r+r_{c}},
\end{equation}
of ref. \cite{Biviano01}, which allows for tangential orbits in the innermost region. 
In \emph{MAMPOSSt} the scale radius for the  $\text{''T''}$ and  $\text{''O''}$ profiles is set to be equal to the scale radius of the mass profile $r_c\equiv r_s$.\\

\section{The simulations of DM halos}
\label{sec:sim}
 We analyze a set of Dark Matter halos extracted from 29 Lagrangian regions of a cosmological $\Lambda$CDM  simulation carried out with the parallel Tree-PM SmoothedParticle-Hydrodynamics
(SPH) code GADGET-3. The parent simulation consists in a periodic box of size $1\,h^{-1}Gpc$ and assumes a flat cosmology with $\Omega_m=0.24$,  $\Omega_{\Lambda}=0.76$, $h=0.72$ and $\sigma_8=0.8$ (see e.g. refs. \cite{Bonafede11,Planellas13,Biffi16}). 
Each Lagrangian region is centered over a massive cluster and re-simulated  with the Zoomed Initial Condition (ZIC) technique of ref. \cite{Tormen97};  particles of mass increasing with distance are used outside the region to correctly reproduce the tidal field on large scales. In the high-resolution region the mass of DM particles is $m_{DM}= 10^9\,h^{-1}M_{\odot}$; the simulation is performed  in such a way that at $ z = 0$ the central halo in not contaminated by low-resolution particles at least out to $5\times r_{200}$. 
Dark Matter particles have a Plummer-equivalent gravitational softening set to $\epsilon=3.75\,h^{-1}\text{kpc}$ in physical units up to $z = 2$ and in comoving units at $z>2$.

Among the central halos, 24 over the 29 in the sample have masses $M_{200}> 8\times10^{14}\,h^{-1}M_{\odot}$, while the remaining 5 are less massive ($M_{200}\sim 1\div 4 \times10^{14}\,h^{-1}M_{\odot}$).

We considered the first and second most massive halos in each Lagrangian region at redshift $z=0$, for a total sample of 58 clusters, requiring that no low resolution particles are included out to $3\times r_{200}$. At $z=0$, the least massive halo is characterized by a virial radius $r_{200}=0.524\,h^{-1}\mpc$  (corresponding to $M_{200}=3.39\times10^{13}\,h^{-1}M_{\odot}$), while the most massive cluster has $r_{200}=2.29\,h^{-1}\mpc$ (corresponding to $M_{200}=2.80\times10^{15}\,h^{-1}M_{\odot}$). The median mass of the sample is $M_{200}=4.82\times10^{14}\,h^{-1}M_{\odot}$.

\begin{figure}
\centering
\includegraphics[width=0.5\textwidth]{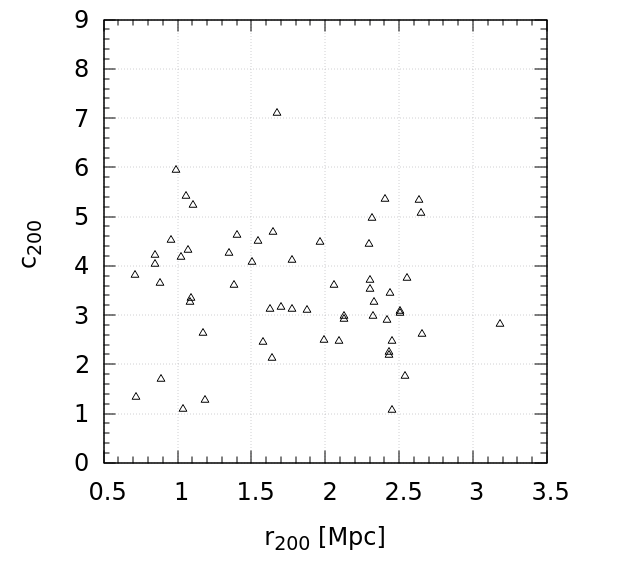}
\caption{\label{fig:fitsim} concentration $c_{200}=r_{200}/r_s$ of the 58 simulated halos in the sample as a function of the radius $r_{200}$. The values of the parameters are obtained by fitting the 3-dimensional  mass distribution with a NFW profile.}
\end{figure}

First, we directly fit the mass profile of each simulated cluster with a NFW model, which is shown to provide a good description for dark matter halos in cosmological simulations (see e.g. refs. \cite{Schaller15,Peirani17})
and adequate fit to GR mass profiles of observed clusters e.g. refs. \cite{Okabe13,Host11}, as already mentioned in Section \ref{sec:theo}. We consider as the center of the cluster the position $(x_C,y_C,z_C)$ of the most gravitationally bound particle. 
In Figure \ref{fig:fitsim} we show the best fit values of the NFW parameter $r_{200}$ and of the concentration $c_{200}=r_{200}/r_s$, obtained by a Maximum Likelihood fit over the total 3-dimensional distribution of particles in every cluster.

We randomly select subsamples of 600 DM particles from each halo in the radial range $[0,1.1\,r_{200}]$ (so that $\sim 530$ particles
are included in the \emph{MG-MAMPOSSt} fit from $0.05\,\mpc$ to $r_{200}$\footnote{As discussed in ref. \cite{Pizzuti17}, generally dynamical relaxation cannot be assumed beyond the virial radius, so the \emph{MAMPOSSt} method is applied only up to $r_{200}$. As for the lower limit, we excluded the innermost region where in real clusters the dynamics is dominated by the brightest central galaxy (BCG, see e.g. ref. \cite{Biviano01})}). In this way we are not considering the problem of interlopers, i.e. galaxies not gravitationally related to the cluster, lying in projection within the range of analysis. The interlopers distribution, if not accurately taken into account, introduces additional spurious effects in the dynamics mass profile reconstruction (see e.g. ref. \cite{mamon10}). In this paper we focus on the possible systematics due to the dynamics of the cluster members only, we will adress the effect on interlopers in forthcoming analyses.
The number of particles within $r_{200}$ has been chosen as a fair compromise between a rich statistics of cluster members and the number of spectroscopic redshift which can be available from present and future surveys.

\begin{figure}
\centering
\includegraphics[width=1.0\textwidth]{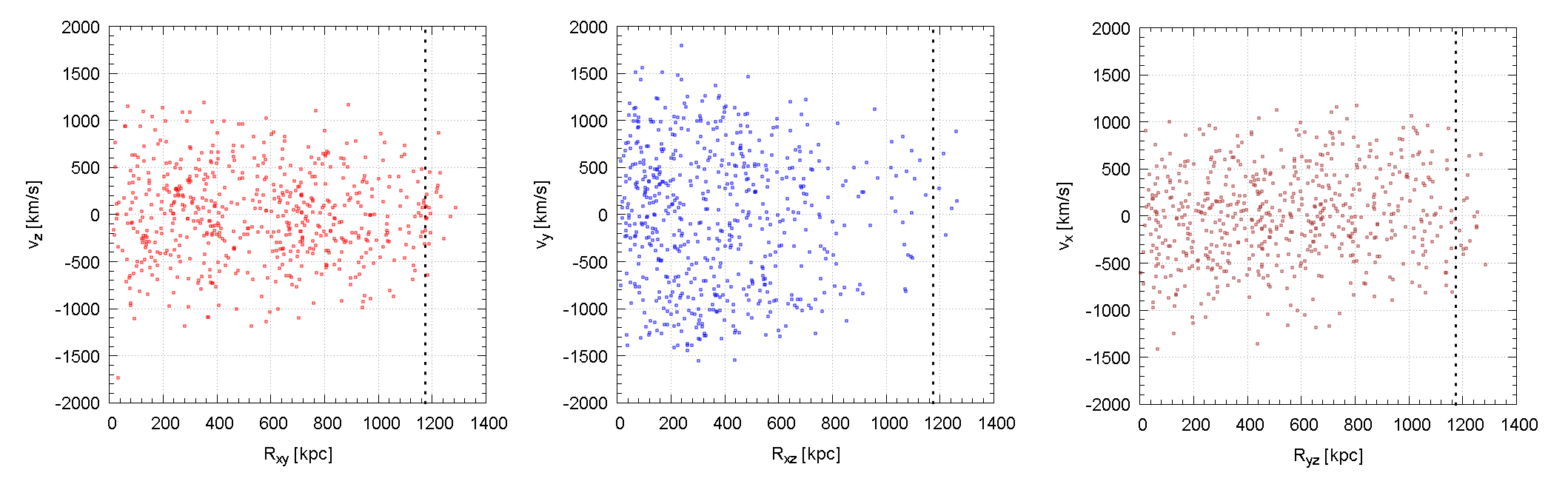}
\caption{\label{fig:phase}Projected phase spaced along the three cartesian axes $\{x_i\}=\{x,y,z\}$ for one subsample of 600 particles derived from the fifth halo in group 0. Each plot shows $R_{i,j}=\sqrt{x_i^2+x_j^2}$ vs $v_{i}$; the black dotted vertical line represent the value $R\equiv r_{200}$. }
\end{figure}

We consider each bi-dimensional projection of each subsample, chosen along the cartesian axes, as an independent phase space; this fact takes into account our ignorance about the orientation with which a generic cluster is observed on the sky. It is also worth to notice that the simulated halos we analyze in this paper contain on average $\gtrsim 10^5$ DM particles within $r_{200}$. This means that a random selection of 600 particles within the cluster range produces realizations of phase spaces which can be very different each other even for the same halo. The completeness of the sample of tracers highly affects the distribution $(R,v_{los})$, giving rise to a bias in the mass profile reconstruction. For this reason, we consider 10 different subsamples for each clusters for a total of 1740 phase spaces  which have been analysed with the \emph{MG-MAMPOSSt} code. As an example, In Figure \ref{fig:phase} we show the three projected phase spaces, in each cartesian direction, derived from one selection of particles for a typical halo in our sample.\\

In linear $f(R)$ gravity, we estimate the probability distribution of the additional degree of freedom $m_{f_{R}}=1/\lambda$ by fitting the data of each phase space with the expression eq. \eqref{eq:phimod} for the gravitational potential $\Phi$, along with the four anisotropy models described Section \ref{sec:mam} in the \emph{MG-MAMPOSSt} procedure.
Since we are interested only in calibrating the effect induced by systematics in our constraints on $m$, in order to avoid the effect of the statistical degeneracy discussed in Appendix \ref{app:const}, we fix the parameters $r_s$ and $r_{200}$ of the mass profile to be equal to the "true" values obtained by the preliminary fit of the 3-dimensional mass distribution. This is equivalent to assume the presence of additional information provided by a probe such as gravitational lensing - which is sensitive to the GR mass profile parameters $r_s,\,r_{200}$ but not to the velocity anisotropy - with infinitely narrow errorbars. This leaves as a free parameters in our analysis $m$ and the velocity anisotropy parameter $\beta_i \equiv \beta_C,\,r_{\beta},\beta_{\infty}$ for the $\text{''C''}$, $\text{''ML''}$ and $\text{''T''/''O''}$ models respectively.

As shown in Appendix \ref{app:const}, for a synthetic halo for which all the assumptions are satisfied (i.e. isolated spherically symmetric distributions of collisionless particles in dynamical equilibrium), \emph{MG-MAMPOSSt} provides results that are always compatible with GR within $68\%$ C.L., independently of the mass of the cluster and the anisotropy model used to generate the object. Therefore, any departure from this condition is a measure of the systematic uncertainties affecting the analysis.

\section{Results}
\label{sec:results}
We run the \emph{MG-MAMPOSSt} code over the full sample of projected phase spaces to fit together the scalaron mass $m$ 
and the velocity anisotropy parameter $\beta_i$, 
adopting 4 different anisotropy models, namely the $\text{''T''}$, $\text{''ML''}$, $\text{''O''}$ and $\text{''C''}$ profiles of eqns. \eqref{eq:betat}, \eqref{eq:betaml},  \eqref{eq:betao} and \eqref{eq:betac} respectively. Looking at the marginalized distribution of $m$, we classify the phase spaces as "Fair"  (F) if the \emph{MAMPOSSt} analysis agrees with GR at 68\% C.L., "semi-Fair" (SF) if GR is included within 95\% C.L. and "not Fair" (NF) if the tension with standard gravity is larger than the $95\%$ limit. The percentage of each class is shown in Figure \ref{fig:perc} for all the $\beta$ profile models assumed in our analysis. As discussed in Appendix \ref{app:const}, we assume that GR is recovered for
$\log(m)\gtrsim 1.8$, which corresponds to a galaxy-scale interaction range $\lambda\le 0.015\,\mpc$.
%

\begin{figure}[ht]
\centering
\includegraphics[width=0.7\textwidth]{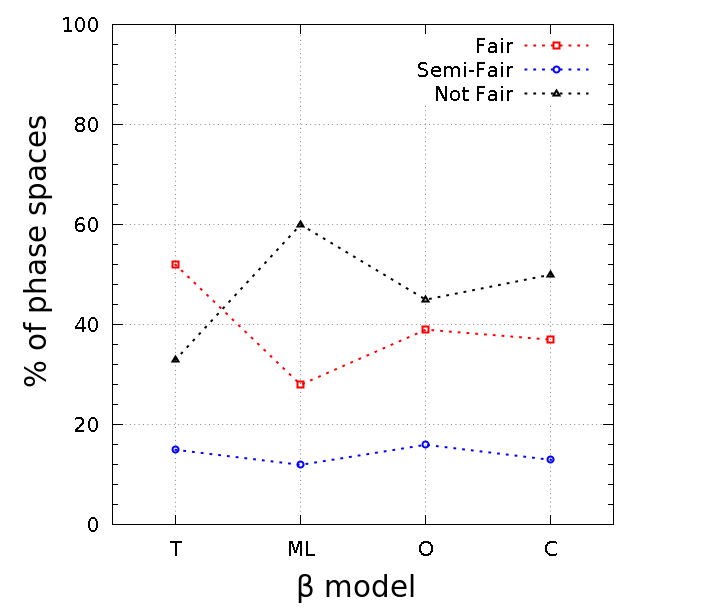}
\caption{\label{fig:perc}Percentage of simulated phase spaces which give a marginalized likelihood of $m$ compatible with GR at $68\%$ C.L. (Fair), at $95\%$ C.L. (Semi-Fair) and excluding GR at more than $95\%$ C.L. (Not Fair) when running \emph{MAMPOSSt} with 4 different anisotropy models. }
\end{figure}

\begin{figure}[ht]
\centering
\includegraphics[width=1.0\textwidth]{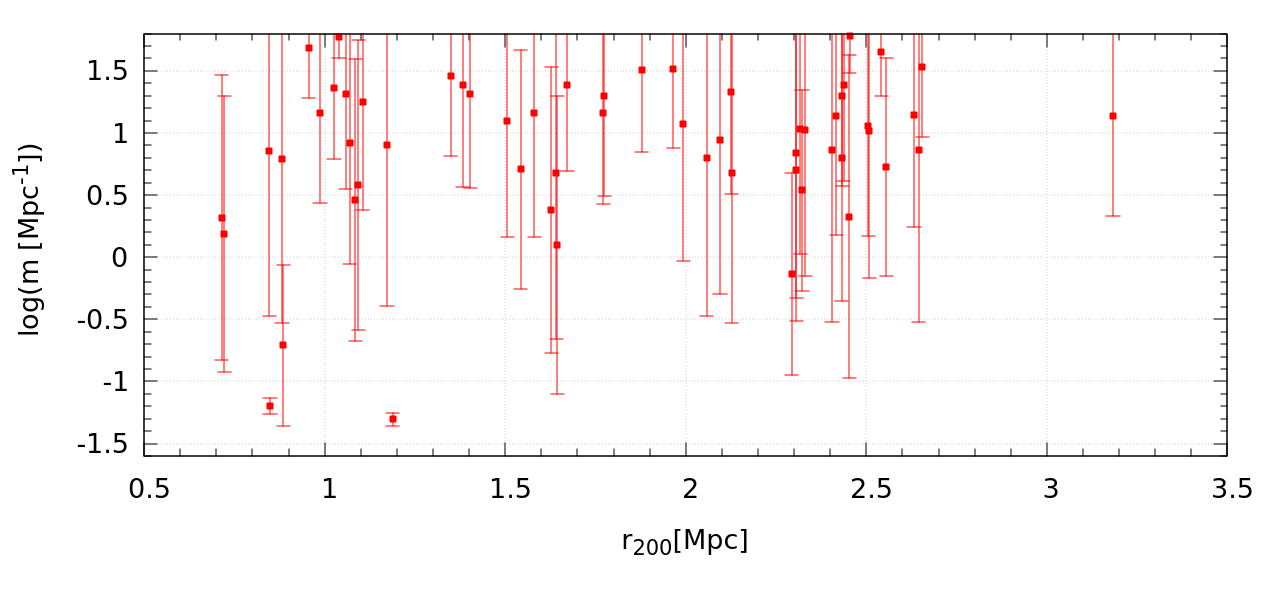}
\caption{\label{fig:mdev}Averaged upper limit on $log(m)$ obtained from the analysis of all the subsamples considered for each simulated cluster, as a function of $r_{200}$. The errorbars indicate the standard deviation, which quantifies the spread of the upper limits found by the \emph{MG-MAMPOSSt} analysis.}
\end{figure}

The results differ when varying the ansatz on $\beta(r)$;
as already mentioned in Section \ref{sec:mam}, changing the model of $\beta$ is an additional source of systematics in the dynamical analysis. 
On average, less than $\sim40\%$ of the sample is suitable to be used as a test for gravity, regardless of the choice of the anisotropy profile.  In Figure \ref{fig:mdev} we show, as a function of $r_{200}$, the $68\%$ C.L. upper limit on $\log(m)$ obtained by averaging over the results from all the realizations of the same halo. The errorbars give a measure of the spread of these values, computed as the standard deviation of the upper limits found by \emph{MG-MAMPOSSt} on all the sampled phase spaces for each cluster.

Our  results shows that a significant fraction of simulated clusters would lead to a spurious detection of deviations from GR. This implies that, in order to use observations of the dynamics of clusters to set reliable constraints on deviations from GR, we need to control the systematic effects related to the violations of the assumption on which the Jeans equation is based. One possible approach is to define a suitable method, based on projected phase-space structure, which select those simulated clusters which show the smallest deviations from GR. 

\begin{figure}
\centering 
\includegraphics[width=1.0\textwidth]{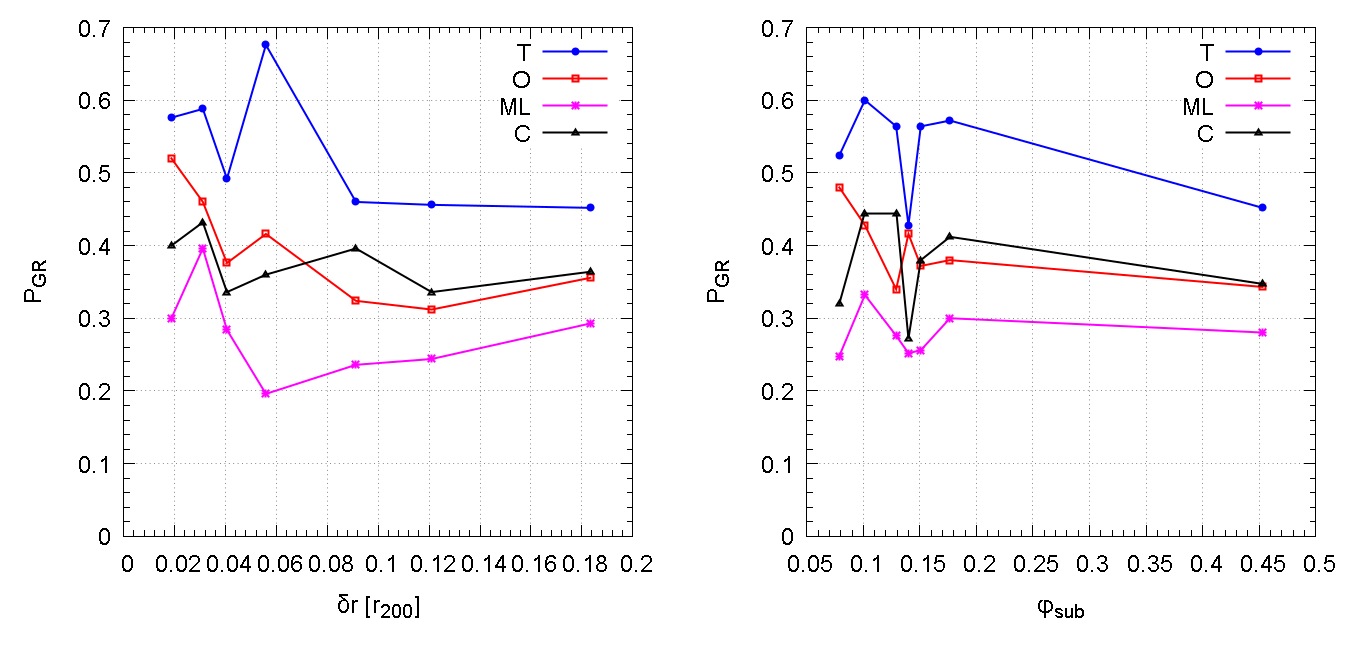}
\includegraphics[width=1.0\textwidth]{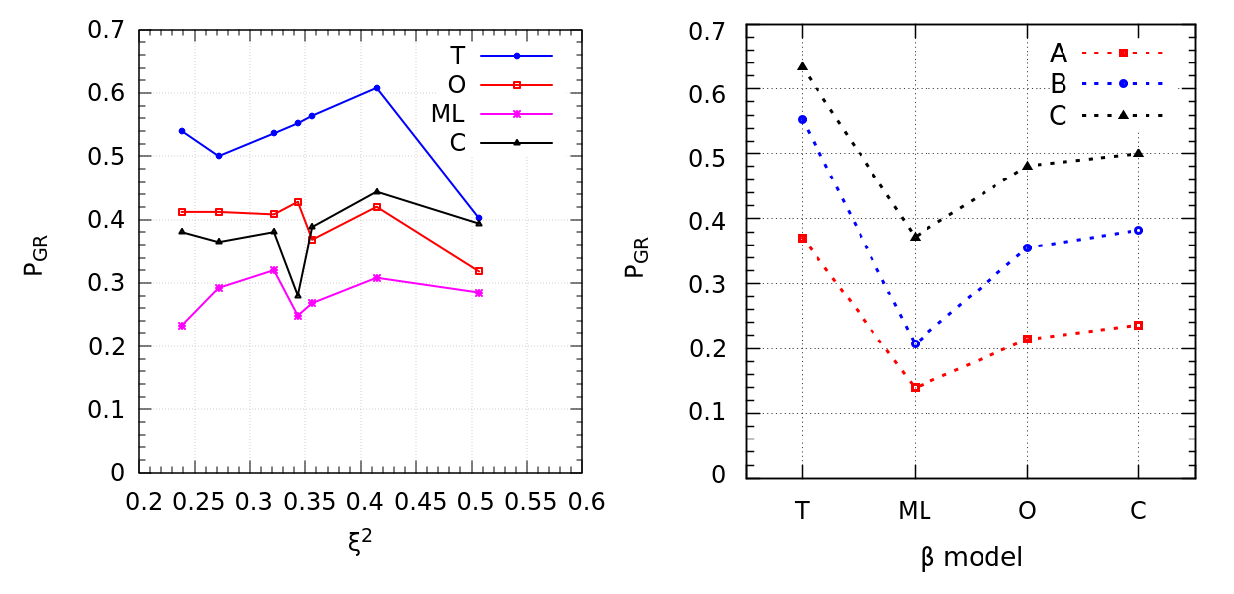}
\caption{\label{fig:theo} $P_{GR}$ as a function of $\delta r$ (left upper plot), $\phi_{sub}$ (right upper plot),
$\xi$(left bottom plot). Different colors in each plot correspond to different models of the velocity anisotropy profile. Bottom right plot: $P_{GR}$ obtained when aligning the l.o.s. with the principal axes of the inertia ellipsoid $A,B,C$ with corresponding semi-axis length $a,b,c$ respectively, displayed as a function of the anisotropy models.}
\end{figure}

In the following, 
we first explore four theoretical criteria, which rely on the knowledge of the 3-dimensional structure of the halo. As such, they cannot be used as a discriminant to select clusters from observational data; however, since they are linked to the dynamical state of the halos, they can help in understanding the effect of the major systematics. The first three criteria are connected to the distribution of particles within the halo and they are listed below:\\
\begin{itemize}
 \item the \emph{center shift} $\delta r$,
 defined as the relative position between the center of mass of the system, computed considering all the particles within the virial radius, and the most gravitationally bound particle, in unit of $r_{200}$:
\begin{equation}
\delta r=\frac{|\mathbf{r}_{CM}-\mathbf{r}_{mb}|}{r_{200}},
\end{equation}
\item the fraction $\phi_{sub}=M_{sub}/M_{200}$ of mass in substructures in the halo with respect to the virial mass of each cluster,
\item the shape of the halo inertia ellipsoid, expressed in terms of the eigenvalues of the inertia tensor:
\begin{equation}
\mathcal{I}_{ij}=\frac{1}{M}\sum_{k}m_kx^{(k)}_ix^{(k)}_j.
\end{equation}
In the above equation, $k$ runs over all the particles within $r_{200}$, while $i,j$ label the three spatial dimensions; $m_k$ is the mass of each particles and $M=\sum_{k}m_k$ the total mass, corresponding to $M_{200}$ in this case.\\
\end{itemize}
The center shift is widely used in numerical simulations to classify the dynamical state of a cluster (see e.g. ref. \cite{Biffi16}), while the eigenvalues of $\mathcal{I}$, $a^2<b^2<c^2$, are related to the length of the semi-axes of the inertia ellipsoid $a,b,c$ (e.g. ref. \cite{Chisari17}), and thus parametrize deviation from spherical symmetry. We define the ratios $\xi=1-a^2/c^2$ and $\zeta=1-b^2/c^2$ such that in the case of a perfect sphere $\xi=\zeta=0$.\\
In Figure \ref{fig:theo} we show the probability $P_{GR}$ of finding a cluster consistent with GR predictions within $68\%$ C.L. as a function of $\delta r$ and $\phi_{sub}$ in the upper left and right panels, and of $\xi$ in the lower left and plots respectively (the behavior of $\zeta$ is the same). Each line refers to a different anisotropy model $\beta(r)$.
No evident effects are observed when considering those criteria. In particular, while a mild dependence of $P_{GR}$ can be spotted for the center shift and fraction of mass in substructures, the eigenvalues of the inertia tensor seem to be totally uncorrelated with the probability of finding clusters in agreement with GR.\\ This shows that the above mentioned parameters are not a good criteria to identify those clusters which are affected by systematics in the recovery of the mass profile and, ultimately, on the constraints on GR deviations. One can think that the lack of correlation between the shape of the halo and $P_{GR}$ could be a consequence of the generic orientation of the halo's inertia principal axes with respect to the line of sight. As already shown by ref. \cite{Mamon01}, the \emph{MAMPOSSt} method better recovers the dynamical parameters in GR when the line of sight is aligned with the direction orthogonal to the major axis of the cluster ellipsoid. This is a consequence of the fact that generally mergers occur along major axes; thus the largest effect on the l.o.s. velocity distribution is found when the observer is aligned with the halo major axis. For this reason, we re-sampled the phase spaces by aligning the line of sight  with the axes of the inertia tensor, instead of with the cartesian coordinate system. Despite the overall percentage of clusters in agreement with GR remains unchanged, a correlation between $P_{GR}$ and the direction of the inertia axes has been observed. More in detail, as shown in the right panel of Figure \ref{fig:theo}, when the line of view is aligned with the major axis "$A$" of the halo, $P_{GR}$  is always lower, independently of the anisotropy model used in the \emph{MAMPOSSt} fit. This confirms that the shape and elongation of clusters are not the main source of systematics for our method, while the direction along which a cluster is observed on the plane of the sky introduce a relevant effect on the constraints obtainable for MG models. 

Determining the angle between a cluster major axis and the line-of-sight is not an easy observational task, albeit feasible when comparing the X-ray emission to the Sunyaev-Zel'dovich signal (see, e.g. ref. \cite{Sereno2006}). In the optical, clusters with their major axis aligned along the line-of-sight might be identified by the circular shape of their BCG and/or of their galaxy spatial distribution.

\begin{figure}
\centering 
\includegraphics[width=0.9\textwidth]{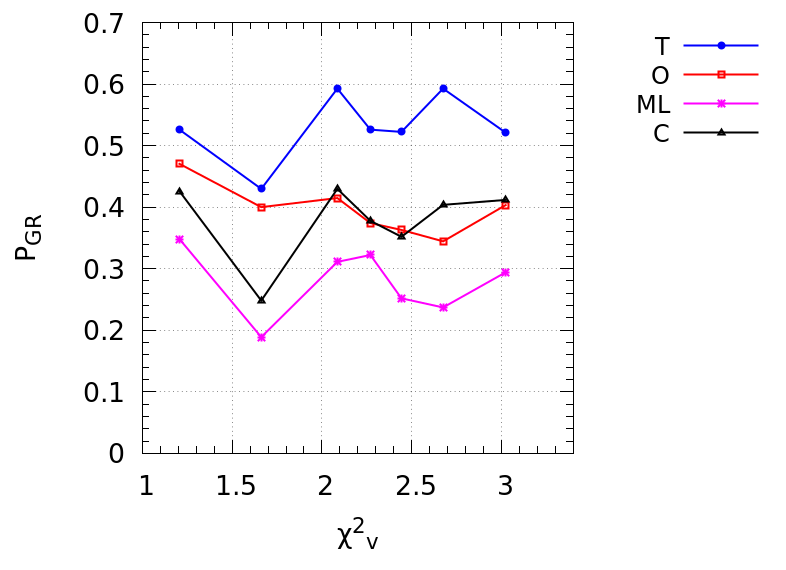}
\caption{\label{fig:theoMB} $P_{GR}$ as a function of the reduced $\chi^2_v$ obtained by fitting the modulus of the velocity of each halo particle with a Maxwell-Boltzmann distribution, eq. \eqref{eq:mb}. Different colors indicate different models of the velocity anisotropy profile.}
\end{figure}

The last theoretical study we perform is related to the 3-dimensional distribution of tracers velocities in the  rest frame of the halo. In particular, as discussed in Section \ref{sec:mam}, the \emph{MAMPOSSt} method assumes a 3D Gaussian probability distribution function, which means that the modulus $v=(v_x^2+v_y^2+v_z^2)^{1/2}$ follows a Maxwell-Boltzmann distribution:
\begin{equation}\label{eq:mb}
\rho(v)=\frac{\mathcal{N}}{a^3}v^2e^{-v^2/(2a^2)}.
\end{equation}
We can assess if deviations from Gaussianity of the velocities 3-dimensional distribution can affect the constraints on the scalaron mass $m$ by performing a Maximum Likelihood fit of eq. \eqref{eq:mb} to the data  and computing the reduced $\chi^2_v=-2\log\mathcal{L}/N_{dof}$, where $\mathcal{L}=\mathcal{L}(v|\mathcal{N},a)$ is the Likelihood and $N_{dof}$ indicates the number of degrees of freedom. In Figure \ref{fig:theoMB} we show the behavior of $P_{GR}$ as a function of $\chi^2_v$ for the four models of the velocity anisotropy.
Again, we are not able to spot any observable effect, confirming the statement of ref. \cite{Mamon01} about the robustness of \emph{MAMPOSSt} with respect to deviations from 3D-Gaussianity, which remains valid also for the \emph{MG-MAMPOSSt} procedure in modified gravity frameworks.  

The results of this analysis suggest that the major spurious effects on constraining departures from GR have to be seeked looking at the structure of the projections, i.e. the projected distribution of particles and the measured l.o.s velocities. \\
As such, we now introduce two  additional observational criteria which are based on the information that can be directly extracted form the projected phase space;  thus, these methods can be used to carry out a "cherry picking" of those clusters which are best suited to set constraints on GR deviations. \\

\begin{figure}
\centering
\includegraphics[width=1.0\textwidth]{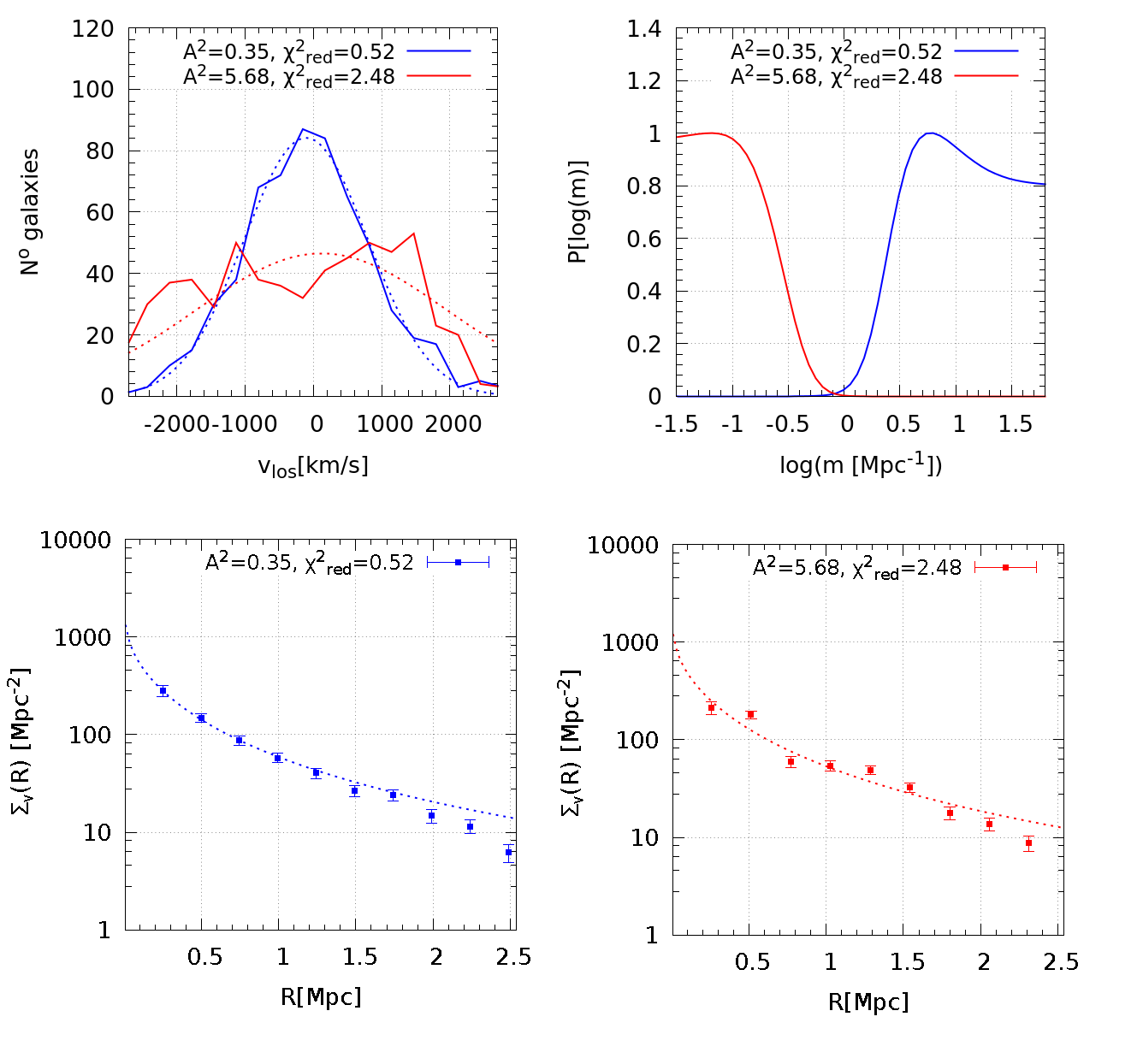}
\caption[L.o.s. velocity distributions and projected density profiles for two values of $A^2, \chi^2_{red}$ and corresponding $P(\log(m))$.]{\label{fig:vlos_A2}Upper left plot: line of sight velocity distributions of two projected phase spaces characterized by a large and a small value of $A^2,\, \chi^2_{red}$ (red and blue solid curves respectively). The dashed lines shows the best fit gaussians in the two cases. Upper right plot: corresponding distributions $P[\log(m)]$ obtained by applying \emph{MG-MAMPOSSt} procedure on these projected phase spaces assuming a "T" velocity anisotropy model in the fit. Lower plots: projected binned number density profiles (points with errorbars) and best fit pNFW profiles (dashed lines) obtained with the method of ref. \cite{Sarazin80}. Left: small values of $A^2,\,\chi^2_{red}$. Right: large values of $A^2,\,\chi^2_{red}$. }
\end{figure}

\textbf{Anderson Darling Coefficient}\\
In analogy to the theoretical analysis, the first criterion is based on quantifying deviations from Gaussianity of the rest frame velocity distribution of the tracers along the line of sight which, as discussed in e.g. ref. \cite{Roberts18}, is a good indicator of the dynamical relaxation state of a galaxy cluster. Indeed, observations and theoretical studies of relaxed systems point out that at equilibrium the l.o.s distribution of galaxies is well described by a Gaussian distribution, while unrelaxed objects show large departures from Gaussianity (see e.g.  refs \cite{Yahil77,Ribeiro13}); furthermore, ref. \cite{Roberts18} found that this indicator highly correlates with other relaxation proxies from X-ray analyses, suggesting that it is a suitable criterion to describe the overall dynamical state of a cluster.\\
Deviations from Gaussianity can be quantified by the so called Anderson-Darling (AD)  test  \cite{anderson1952}, which determines how different are the cumulative distribution functions of the data set and of the ideal Gaussian case. The AD coefficient $A^2$ is defined as:
\begin{equation}\label{eq:ad}
A^2=-n-\frac{1}{n}\sum_{i=1}^{n}\left\{\left[\log \Phi(x_i)+\log(1- \Phi(x_{n+1-i})) \right](2i-1)\right\}.
\end{equation}
In the above equation $x_i$ is the $i^{th}$-element of the data set, in ascending order, which in our case corresponds to the velocity of the $i^{th}$ particle; $\Phi(x_i)$ is the value of the cumulative Gaussian distribution function\footnote{The AD statistics can be used to test other known distributions, such as flat or exponential, changing the form of $\Phi(x)$} at $x_i$. A large value of $A^2$ means large deviation from Gaussianity. As an example, the upper left panel of Figure \ref{fig:vlos_A2} illustrates the shape of the l.o.s. velocity distribution for two projected phase spaces in our sample characterized by a value of $A^2=5.68$ (red curve) and $A^2=0.35$ (blue curve); on the upper right panel we show the corresponding marginalized distributions $P[\log(m)]$ from the \emph{MG-MAMPOSSt} analysis with a Tiret model for $\beta(r)$. As we can see, while the phase space with $A^2<1$ produce a constrain on $m$ in agreement with GR, the analysis of the projected phase space with large $A^2$ excludes standard gravity at more than $4\sigma$.  \\

\textbf{2D Number density Chi-square}\\
The second method that we considered in this analysis is based on the reduced $\chi^2$ resulting from fitting the projected number density  profile of galaxies $\Sigma(R)$  with a projected NFW (pNFW) density profile. The value of $\chi_{red}^2$ incorporates the effect of several systematics, such as departures from spherical symmetry, the presence of substructures and the uncertainties in the choice of the parametrization of the number density profile. Indeed, although it is true that the NFW model provides excellent agreements both in simulations and observations, as mentioned before other parametrizations, characterized by a larger number of degrees of freedom, could better describe the mass profile of DM halos (see e.g. ref. \cite{Retana12} and references therein) especially when the cluster exhibits strongly unrelaxed features.\\
The fit has been performed by a Maximum Likelihood approach of ref. \cite{Sarazin80} (see also e.g. ref. \cite{mamon10}) which does not require binning of the data. In particular, given $\rho(R_k|r_\nu,\Sigma_{bg})$ the probability to find a galaxy at a projected position $R_k$ in a spherical NFW model with scale $r_{\nu}$ and projected background density $\Sigma_{bg}$ (see eq. (1) in ref. \cite{Sarazin80}), the reduced $\chi^2$ is defined as 
\begin{equation}
\chi_{red}^2=-\log[\mathcal{L}_{\nu}(r_{\nu},\Sigma_{bg})]/N_{dof},
\end{equation}
where, as above, $N_{dof}$ is the number of degrees of freedom in the fit and
\begin{displaymath}
-\log[\mathcal{L}_{\nu}(r_{\nu},\Sigma_{bg})]=-\sum_{k}\log[\rho(R_k|r_\nu,\Sigma_{bg})].
\end{displaymath}
In our case we set the background density to zero in the fit as we are considering only particles within a sphere of radius $1.1\,r_{200}$ from the cluster center.  We show in the lower plots of Figure \ref{fig:vlos_A2} the binned projected number density profiles, corresponding to the two above mentioned phase spaces, as the square points. The dashed lines indicate the best fit pNFW model obtained by the Maximum Likelihood approach. Note that the phase space with the lower $\chi^2_{red}=0.52$ (left plot) corresponds to the distribution $P[\log(m)]$ in agreement with GR, while the phase space  characterized by  $\chi^2_{red}=2.58$ is also the one where GR is excluded by the \emph{MG-MAMPOSSt} analysis.
\begin{figure}
\centering
\includegraphics[width=1.0\textwidth]{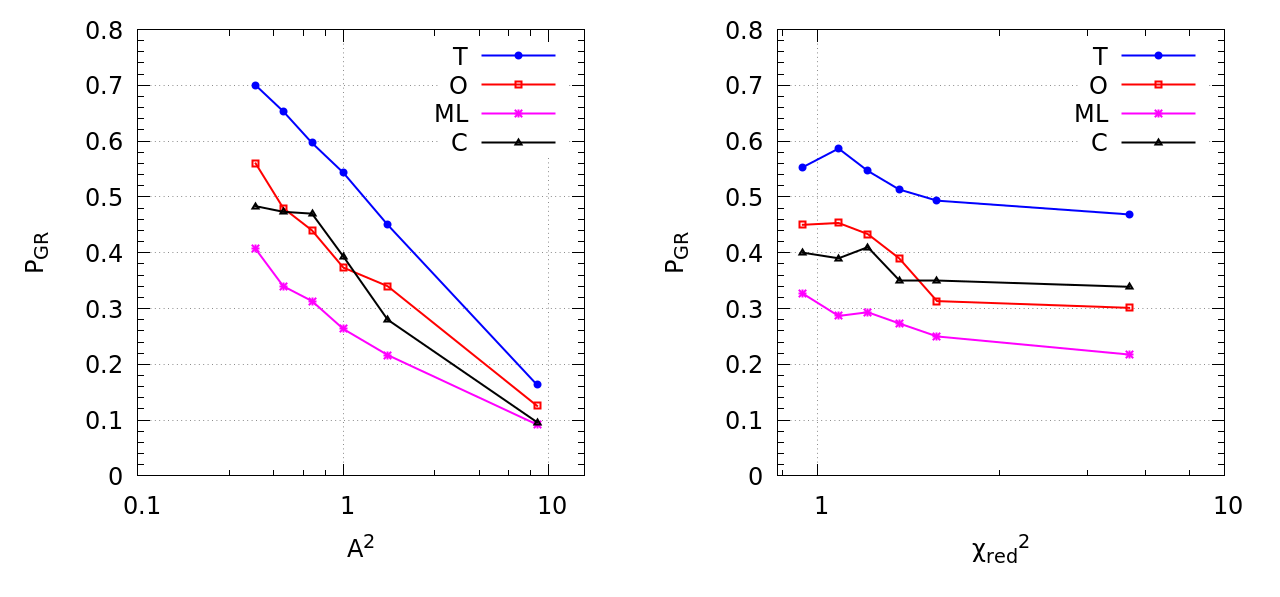}
\caption[$P_{GR}$ as a function of $A^2$ (left) and $\chi^2_{red}$ (right).]{\label{fig:chiads_all}Probability that, given a generic phase space, the \emph{MG-MAMPOSSt} analysis produces results compatible with GR predictions within $1 \sigma$,  expressed as a function of the AD coefficient $A^2$ (left plot) and the reduced $\chi^2_{red}$ (right plot). Different colors and point types refer to different anisotropy models. The binning is computed such that each bin contains the same number of clusters.}
\end{figure}

\begin{figure}[ht]
\centering
\includegraphics[width=0.70\textwidth]{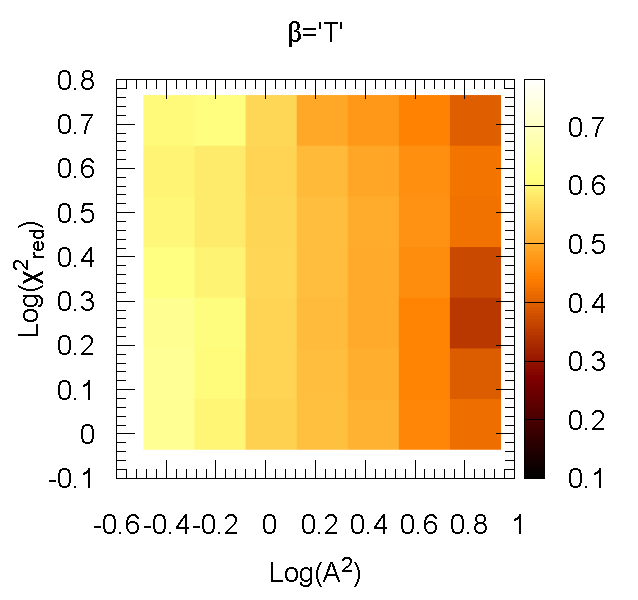}
\caption[Color map of $P_{GR}$ as a joint function of $A^2$ and  $\chi^2_{red}$.]{\label{fig:prob_dual}Values of probability to be consistent with GR within $68\%$ C.L., $P_{GR}$, for different values of $A^2$ and $\chi_{red}^2$ 
parameters in the case of the "T" anisotropy model. Each $i^{th}$ bin is colored according to the probability of identifying a phase space for which consistency with GR is found in the range of values $[A^2_{(i-1)},A^2_{(i)}]$ and 
$[{\chi^2_{red}}_{(i-1)},\,{\chi^2_{red}}_{(i)}]$  .}
\end{figure}

In Figure \ref{fig:chiads_all} we plot the probability $P_{GR}$   as a function of $A^2$ (left plot) and of  $\chi^2_{red}$ (right plot) for the 4 models of $\beta(r)$. Each bin has been built with the same number of phase spaces. Interestingly, we found an overall increase of $P_{GR}$ towards lower values of both parameters, with a stronger effect for the AD statistics. We checked that the result does not depend on the chosen model of the velocity anisotropy parameter.  This means that phase spaces suitable for the application of our method to constrain GR deviations should be identified among those clusters characterized by an almost Gaussian l.o.s. velocity distribution ($A^2<1$) and a reduced chi square in the fit of the projected number density profile $\chi^2_{red}<0.5$.\\
To further highlight this behavior, Fig. \ref{fig:prob_dual} shows the variation of $P_{GR}$ for a grid of values of  $A^2$ and $\chi^2_{red}$  in the case of the "T" anisotropy profile.   A similar behaviour has been found for the other anisotropy models. As expected, the probability rises in the region corresponding to the lower values of chi square and AD coefficient; in particular, for $A^2\lesssim 1$  and $\chi^2_{red} \lesssim 0.5$ it  reaches $\sim 80\%$ in the case of "T" model and $\sim 43\%$ in the case of "ML" model, increasing more than $70\%$ with respect to the average values. 

\subsection{Chameleon $f(R)$ gravity}
 \label{sec:screen}
\begin{figure}[ht]
\centering
\includegraphics[width=0.70\textwidth]{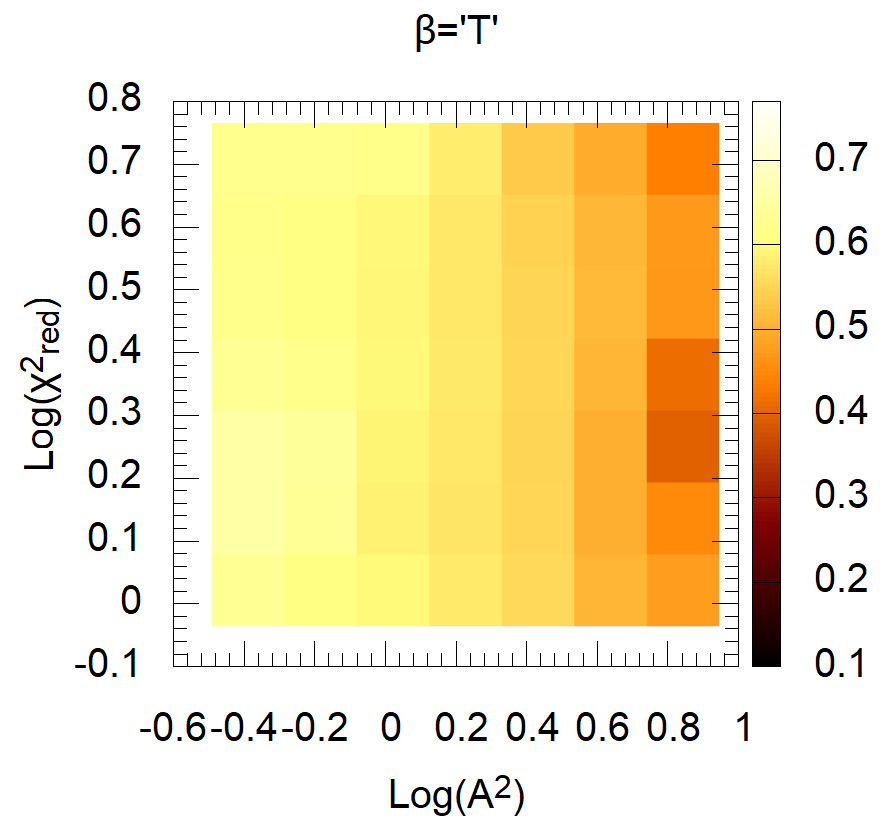}
\caption[Color map of $P_{GR}$ as a joint function of $A^2$ and  $\chi^2_{red}$.]{\label{fig:prob_dualSc}The same as Figure \ref{fig:prob_dual} but obtained considering a simplified treatment for the screening mechanism in chameleon $f(R)$ gravity.}
\end{figure}

.
As already mentioned in Section \ref{sec:introd}, viable models of $f(R)$ gravity at cosmological scales are characterized by a suppression of the additional degree of freedom in high density regions. In particular, an environment-dependent field is used to recover GR at small scales; this procedure is known as the chameleon mechanism (ref. \cite{K2004}). The details of the screening, as well as the full scalaron field profile depend  on the functional form of $f(R)$ and can be derived by solving the non-linear equation of motion (see e.g. ref. \cite{Zhao11,Lombriser17}):
\begin{equation}
\Box f_{,R}=\frac{1}{3}\left[R-f_{,R}R+2f(R)-{8\pi G}\rho_{m}\right].\label{eq:motion}
\end{equation}
Discussing the solution of eq. \eqref{eq:motion} is beyond the purposes of this paper; however, in order to determine qualitatively how the impact of systematics changes when a screening mechanism is implemented in the model, we repeat the analysis assuming a simplified setup for the screening effect. We follow the prescription of ref. \cite{Pizzuti17} which considered the analytical approximation of ref. \cite{Lomb12} to model the chameleon field. As for the functional form of $f(R)$, we consider the popular Hu \& Sawicki parametrization (ref. \cite{Hu2007}):
\begin{equation}
 f(R) \simeq 6\Omega_{\Lambda}-\frac{f_{,R0}}{n}\frac{R_{0,b}^{n+1}}{R^n},
\end{equation}
where the dependence on the Ricci scalar is expressed by a power-law. $R_{0,b}$ and ${f_{,R0}}$ represent the background values computed at $z=0$ for the Ricci scalar and the scalaron field respectively; we choose the exponent $n=1$.
The approximation consists in assuming an instantaneous transition between the linear and the full-screened regime; the radius at which the transition occurs, dubbed as the screening radius, depends on the environmental density and on the background value of the scalaron mass $\bar{m}_{f_R}$. Despite its simplicity, this analytic approach has been shown to correctly reproduce the global behavior of the scalar field; for a more detailed discussion on the procedure, see ref. \cite{Lomb12}. 
As already pointed out in ref. \cite{Pizzuti17}, in the presence of a chameleon mechanism the bounds we can obtain on $\bar{m}_{f_R}$ are much wider than the simple linear case, since for the majority of viable $f(R)$ models a galaxy cluster should be almost entirely screened if $\bar{m}_{f_R}\gtrsim 0.1\, \mpc^{-1}$ (corresponding to $|f_{,R0}|\lesssim 10^{-6}$). We will discuss possible constraints on chameleon MG theories obtainable from future surveys by combining lensing and dynamics cluster mass profiles in a forthcoming paper.
The results of the screening analysis are shown in Figure \ref{fig:prob_dualSc} in the case of "T" anisotropy model (analog for the other $\beta(r)$ profiles). The trend of $P_{GR}$ is similar to what obtained in linear $f(R)$, confirming that the main conclusions of this work remain valid also in the non-linear framework. Nevertheless, it is worth to notice that an overall slightly larger ($5$ to $10\%$) percentage of cluster in agreement with GR has been found in this case when no selection criteria are considered; moreover, the correlation between $P_{GR}$, $A^2$ and $\chi^2_{red}$ appears to be somewhat weaker when the chameleon mechanism is included. This is not surprising, as in the non-linear case only large values of $\bar{m}_{f_R}\ll 1$ 
 produce significant effects on the cluster dynamics. For smaller values, the screening mechanism turns immediately very efficient: field gradients are suppressed and the motion of cluster galaxies become identical to GR. Thus, too small deviations from dynamical relaxation and spherical symmetry, which can still be a relevant source of systematics in linear $f(R)$, cannot now be compensated by the presence of a spurious intermediate-scale interaction range $\lambda=1/\bar{m}_{f_R}\sim 1 \,\mpc$. For this values the halo is totally screened with the exception of a thin shell at the cluster's edge; as such, GR and MG scenarios are almost indistinguishable.

\section{Discussions and conclusions}
\label{sec:conc}
In this paper we have examined a set of $\Lambda$CDM simulated cluster-size Dark Matter halos  in order to estimate the impact of systematics in constraining modification of gravity using a combination of dynamics and lensing cluster mass profile determinations, specifically focusing on the dynamics of cluster member galaxies. 

As a case of study,  we have considered the framework of linear and screened $f(R)$ gravity, where deviations from GR  are expressed in terms of the mass of the additional degree of freedom, $m\equiv m_ {f_R}$. We applied the \emph{MG-MAMPOSSt} method (refs. \cite{Mamon01,Pizzuti17}) to reconstruct the cluster mass profile in modified gravity models with a Maximum Likelihood approach, by solving the spherical Jeans' equation in the projected phase space ($R,v_{los}$) of member galaxies. The method assumes dynamical relaxation and spherical symmetry of the total matter distribution.
In order to understand how much the breakdown in the assumptions of the analysis affects the constraints on $m$, we have analyzed a sample of 1740 projected phase spaces of galaxy clusters extracted from cosmological N-body simulations carried out with with the GADGET-3 code. These clusters are taken at $z=0$ and have masses in the range $3\times10^{13}\,h^{-1}M_{\odot}\div 4\times 10^{15}\,h^{-1}M_{\odot}$; for each halo 10 projected phase spaces have been considered by randomly sampling 600 
Dark Matter particles in the radial range $[0.05\,\mpc,1.1\,r_{200}]$ . The results show that $\sim 60\%$ of clusters in a $\Lambda$CDM Universe (where GR is assumed) produce a spurious detection of modified gravity when no selection criteria are used. This illustrates that the impact of systematics in the proposed method, in particular deviations from spherical symmetry and departures from relaxation of the cluster, plays a dominant role; an accurate control and calibration of such effects is thus required in order to claim our procedure robust. \\

For this reason, in order to identify a sample of galaxy clusters to be used for constraining deviations from GR, we have defined two observational parameters which correlate with the probability to find clusters in agreement with GR predictions and which can help in identifying the suitable clusters for the application of our method. These parameter are  the Anderson-Darling coefficient $A^2$ of the l.o.s. velocity distribution and the reduced chi square $\chi_{red}^2$ of the projected number density profile of the tracers, which are connected to the main systematics affecting the analysis and can be directly measured  from the projected phase space. The first one identifies deviation from Gaussianity of the l.o.s. velocity distribution, which is connected to the lack of dynamical relaxation, while the second criterion is related to the capability of a smooth NFW density  profile to describe the projected distribution of tracers. The analysis we carried out shows that spurious detections of $\lambda\equiv 1/m >0$ in the marginalized distribution are correlated with the values of these two observational criteria; in particular, the probability to find spurious detection of modified gravity decreases to $\sim20\%$ by selecting clusters that have $A^2 < 1$ and $\chi^2_{\nu}< 0.5$. Moreover, the trend is independent of the parametrization of the velocity anisotropy $\beta(r)$ in the \emph{MAMPOSSt} analysis, which is the major source of uncertainties in the dynamics mass profile reconstruction.  This demonstrates that the proposed criteria can be used to identify the suitable clusters for the application of our method to constrain deviations from GR based on the comparison of mass profiles of clusters reconstructed from dynamics of member galaxies and from lensing observations. \\

The joint lensing and dynamics analysis in linear $f(R)$ gravity of ref. \cite{Pizzuti17} on the data of two CLASH clusters MACS 1206 and RXJ 2248, indicates agreement with GR in the case of MACS 1206, while a slight preference of $\lambda>0$ has been found for the cluster RXJ 2248. In particular, at $\Delta\chi^2=2.71$ they obtain $\lambda_{1206}\le 1.61$ (345 galaxy cluster members available in the \emph{MG-MAMPOSSt} fit), $\lambda_{2248}\ge 0.14$ (981 members available in the \emph{MG-MAMPOSSt} fit) respectively. If we apply the above mentioned criteria over the phase spaces of these two clusters we find that:
\begin{displaymath}
    A^2_{1206}=0.99\,\,\,\,\,\,\,\,\, \left(\chi^2_{red}\right)_{1206}=0.87\,,
\end{displaymath}
\begin{equation}
    A^2_{2248}=2.77\,\,\,\,\,\,\,\,\, \left(\chi^2_{red}\right)_{2248}=1.58\,.
\end{equation}
Thus, the phase space of MACS 1206 exhibits low values of the observational parameters, as expected by clusters consistent with GR according to the study performed in this paper.  On the other hand RXJ 2248 is characterized by large $A^2,\chi^2_{red}$, suggesting that the deviation from GR prediction found from the joint lensing+dynamic analysis can be actually sourced by systematic effects. It is worth to point out the RXJ 2248 is an overall relaxed cluster (see ref. \cite{Pizzuti17} and references therein) as the dynamics and lensing mass profiles obtained in GR are in agreement within $68\%$ C.L. However, it is worth to remind that in a modified gravity scenario additional degrees of freedom appear and it is not possible to determine \emph{a priori} whether or not tiny departures from the main assumptions, which produces negligible effects in the reconstruction of mass profiles under the assumption of GR, could induce a large systematic impact on the recovery of the parameters describing a modified gravity model.
%

The analysis presented in this paper has important implications for the study of
galaxy clusters to test the nature of gravity over cosmological scales; with the next generation
surveys such as Euclid, LSST, high-quality imaging data over broad area of the
sky will allow lensing mass profile reconstructions to be carried out for
several hundreds clusters in quite large sky region. Moreover,
next generation high-multiplexing spectrographs on 8-meter class telescopes will allow to carry out detailed dynamical studies of a fraction of these clusters, thus providing a large statistical basis for the application of the method presented here. In this context, it is important to devise some criteria, like those considered here, to identify the best-suited clusters for constraining GR deviations from the comparison of lensing and dynamical analyses.  

Several developments of the work presented in this paper can be made in order to better quantify the effects of the above mentioned systematics. It is worth to remind that the results of this analysis have been obtained by studying only the dynamics of particles in simulated Dark Matter halos; the behavior of galaxies in real clusters can be significantly different from that of DM particles, in particular in the innermost regions where the effects of astrophysical processes become relevant (e.g. refs. \cite{Schaller15b,Munari13}). It is important to investigate how the criteria introduced by our study are affected by the presence of baryons (gas and stars) and when considering the dynamics of substructures in high resolution simulations, which better reproduce the dynamics of galaxies in clusters with respect to DM particles. In particular, baryonic processes such as Supernovae feedback or radiative cooling modify the concentration $c_{200}$ of the derived mass profile, flattening it with respect to a NFW towards the cluster center, and produce a bias in the mass reconstructions (see e.g. refs. \cite{Shirasaki18,DelPopolo19});  this should reflect on the constraints on modified gravity parameters, further influencing the relations highlighted in this paper. Specifically, the selection effect based on the reduced $\chi^2$ parameter, which in our case relies on the NFW model assumption, is expected to be larger when baryons are taken into account.
 
 Furthermore, the inclusion of full lensing mass reconstructions in our analysis as well as the study of the effect induced by the presence of interlopers in dynamic mass profile determinations will help in obtaining realistic constraints on modified gravity parameters, at different redshifts, to be compared with real data. \\

%
%



\noindent \textbf{Acknowledgements.} 
LP is partially supported by a 2019 "Research and Education" grant from Fondazione CRT. The OAVdA is managed by the Fondazione Clément Fillietroz-ONLUS, which is supported by the Regional Government of the Aosta Valley, the Town Municipality of Nus and the "Unité des Communes valdôtaines Mont-Émilius. SB acknowledges financial support from
PRIN-MIUR 2015W7KAWC,
and the INFN INDARK grant. BS  is supported by the "FARE-MIUR grant 'ClustersXEuclid' R165SBKTMA"  and partially by "PRIN-MIUR2017 WSCC32 'Zooming into dark matter and proto-galaxies with massive lensing clusters'".


\bibliographystyle{JHEP}
\bibliography{master,master2}

\appendix
\section{Analysis of synthetic Dark Matter halos}\label{app:const}
\begin{figure}
\centering
\includegraphics[width=1.0\textwidth]{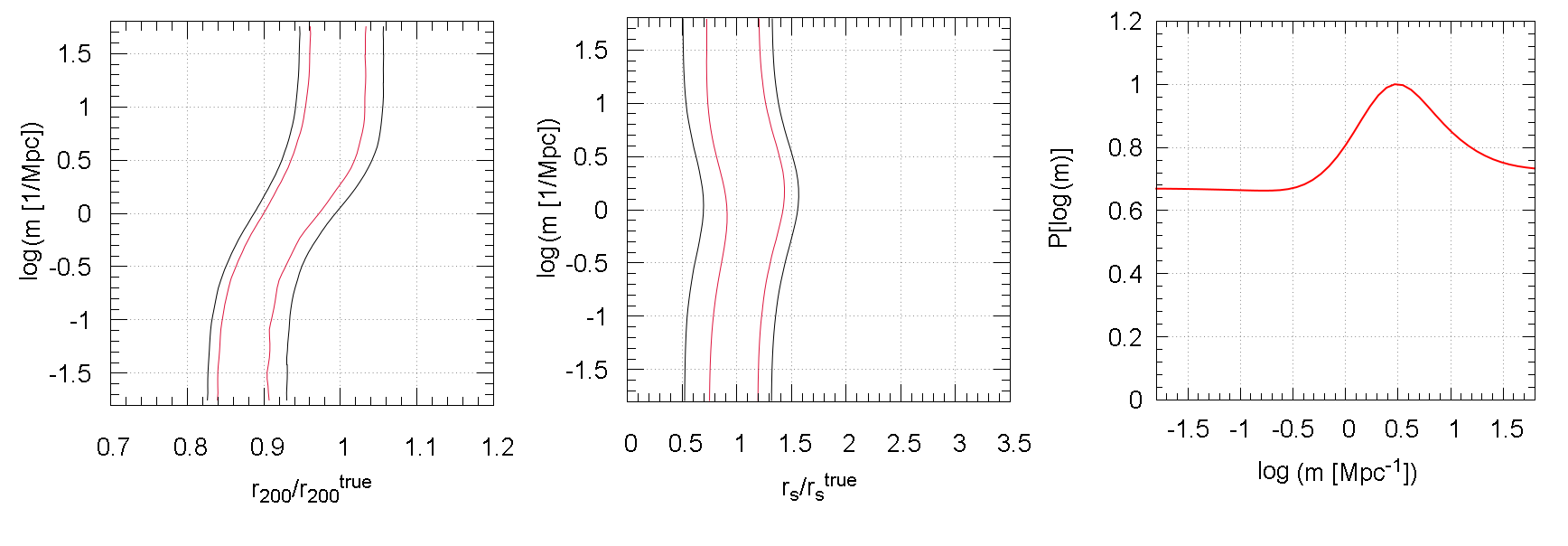}
\caption[2-dimensional distributions in the plane  $(r_{200},\log(m))$, $(r_{s},\log(m))$ and marginalized distribution $P(\log(m))$ from the analysis of one synthetic cluster.]{\label{fig:marg_1815d} 2-dimensional likelihood distributions in the plane  $(r_{200},\log(m))$ (left plot), $(r_{s},\log(m))$ (central plot) and the marginalized distribution $P[\log(m)]$ obtained by the \emph{MG-MAMPOSSt} analysis of one synthetic halo in the sample. The red contours in the left and central plots refer to points at $\Delta\chi^2=2.3$, while the black contours indicate points at $\Delta\chi^2=4.6$.}
\end{figure}

\begin{figure}
\centering
\includegraphics[width=1.0\textwidth]{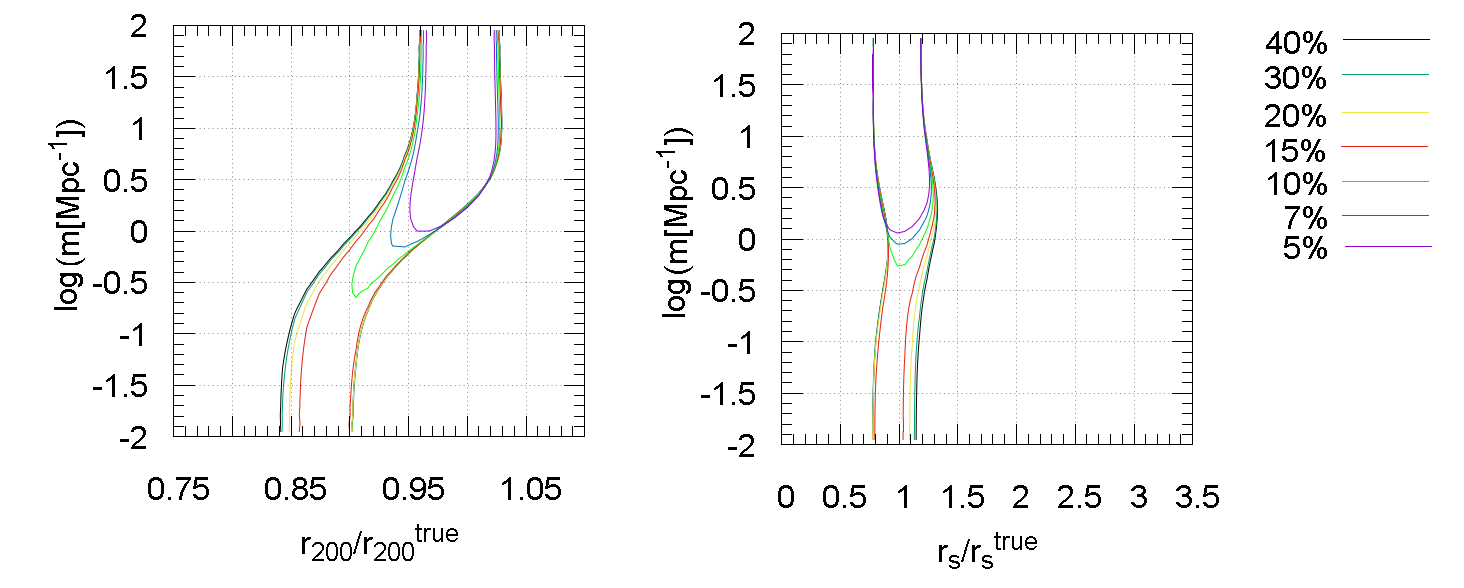}
\caption[2d contours  $(r_{200},\log(m))$ (left) and $(r_s,\log(m))$ (right) for different values of $\sigma_{r_{200}}$ in $P_L(r_s,r_{200})$.]{\label{fig:speriamo} 2-dimensional contours at $\Delta \chi^2=2.3$ in the parameter space  $(r_{200},\log(m))$ (left) and $(r_s,\log(m))$ (right) for different values of $\sigma_{r_{200}}$ in $P_L(r_s,r_{200})$, from $40\%$ (black curves) to $5\%$ (purple curves).}
\end{figure}

\begin{figure}
\centering
\includegraphics[width=1.0\textwidth]{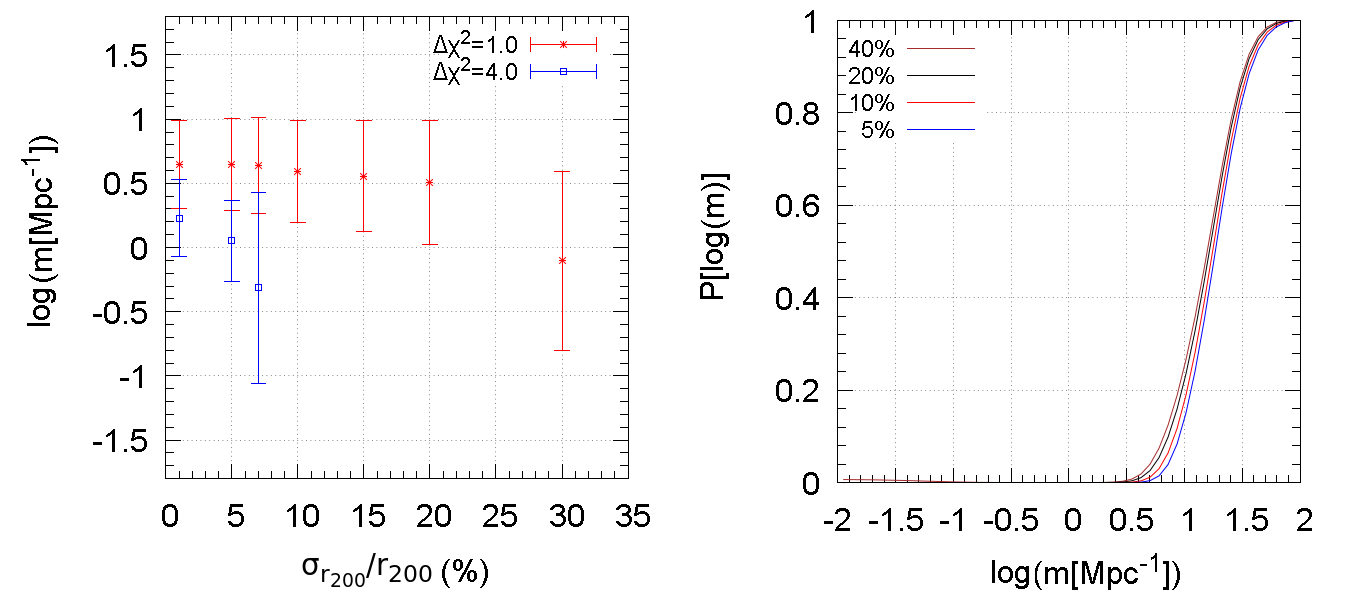}
\caption[Averaged lower limit on $m$ as a function of $\sigma_{r_{200}}$ (left), all-halos-combined distribution of $m$ (right).]{\label{fig:lim_lensm} Left: lower limit on $m$  obtained averaging the lower limits of each marginalized distribution from the analysis of the 15 synthetic halos as a function of $\sigma_{r_{200}}$ in $P_L(r_s,r_{200})$ (points); the error bars indicate the scatter around the mean value computed over this ensemble of synthetic halos. Red: $\Delta \chi^2=1.0$. blue: $\Delta \chi^2=4.0$. Right: all-halos-combined distribution of $m$. Different lines correspond to different values of the prior in $r_{200}$}
\end{figure}

In order to investigate how possible systematics can affect constraints on modified gravity models obtained using a combination of lensing and kinematic information, one needs first to test the reliability of our method in the case when all the assumptions (i.e. spherical symmetry and dynamical relaxation) are perfectly satisfied.   Adopting a $\Lambda$CDM background with $H_0=70\, \text{km}\,\mpc^{-1}s^{-1}$, $\Omega_m=0.3$ and $\Omega_{\Lambda}=0.7$ (where the cosmology enters only in the definition of the virial mass $M_{200}$), we generate a sample of 15 spherically-symmetric, isolated halos made of collisionless particles distributed  as a NFW mass profile. We follow the method described in Section 4.1 of ref. \cite{Pizzuti19},  populating each halo out to $\sim9\,r_{200}$, and assuming the same prescriptions for the parameters $r_s,r_{200}$. We impose to have 1000 particles within the virial radius\footnote{As discussed in ref. \cite{Pizzuti19}, the choice of 1000 particles within $r_{200}$ is an optimistic expectation  of the number of cluster galaxies whose spectroscopic redshift can be measured from present and future surveys.}. The velocity components of each particles at a radius $r$ from the cluster center are assigned by a Gaussian random around zero (i.e. we work in the rest frame of the halo) where the variance $\bold{\sigma}^2(r)$ is given by the solution of the spherical Jeans' equation eq. \eqref{eq:Jeans} with an assumed model for $\beta(r)$. As in ref.  \cite{Pizzuti19}, we impose the Tiret model, eq. \eqref{eq:betat}, with the parameter $\beta_{\infty}$ fixed to $0.5$.\\
The input values of $r_{200}$ and $r_s$ are given in the second and third columns of Table \ref{tab:perfc}, ordered by increasing mass. As discussed in Section \ref{sec:mam},  the version of \emph{MG-MAMPOSSt} code we are adopting requires a parametric model of the projected number density profile of the tracers\footnote{It is important to notice that in general the \emph{MAMPOSSt} procedure does not necessarily involve a parametrization for the physical quantities in the Jeans' equation; nevertheless, for the purpose of this work a parametric approach is worthed to investigate specific class modified gravity models}, with a characteristic scale radius $r_{\nu}$; since we are working with collisionless particles, by construction the number density profile $\nu(r)$ scales exactly as the NFW mass profile $\rho(r)$ (i.e. $r_{\nu}\equiv r_s$). However, the \emph{projected} number density profile, needed to compute the Likelihood, is obtained in \emph{MAMPOSSt} by integrating the 3-dimensional profile along the line of sight, assuming that it extends to infinity. This leads to a value of $r_{\nu}$ which can be slightly different from $r_s$; for this reason we fit the projected number density profile from the phase space of each clusters and we use the best fit values of $r_{\nu}$ as the input for the \emph{MG-MAMPOSSt} analysis (see the fourth column of Table \ref{tab:perfc}). 
\begin{table}
\centering
\begin{tabular}{|c|c|c|c|}
\hline
  ID & $r_{200}\,[Mpc]$ & $r_s\,[Mpc]$ & $r_{\nu}\,[Mpc]$ \\\hline
\hline
1&	0.750&	0.096&	0.086		\\
      
2&	0.900&	0.115&	0.110		 \\
    
3&	1.050&	0.126&	0.128		 \\
    
4&	1.200&	0.200&	0.195		 \\
   
5&	1.350&	0.106&	0.096		 \\
    
6&	1.500&	0.298&	0.269		\\
    
7&	1.650&	0.379&	0.394	 \\
   
8&	1.800&	0.519&	0.493		 \\
   
9&	1.950&	0.448&	0.430		 \\
  
10&	2.100&	0.537&	0.561		  \\
   
11&	2.250&	0.536&	0.483		\\
   
12&	2.400&	0.591&	0.530		   \\
   
13&	2.550&	0.990&	1.035	 \\
    
14&	2.700&	0.957&	1.007	 \\
    
15&	2.850&	1.077&	0.950	 \\
\hline
\end{tabular}
\caption[Parameters of the synthetic halos used in the analysis]{\label{tab:perfc} Input values of the synthetic halos. Second and third column: NFW parameters $r_{200},\,r_s$ for the sample of synthetic halos used in our analysis. The fourth column shows the values of the scale radius $r_{\nu}$ of the projected number density profile of the particles fitted from the phase space, which is additionally required in the \emph{MAMPOSSt} procedure. }
\end{table}
 We apply the \emph{MG-MAMPOSSt} method to the synthetic clusters sample in linear $f(R)$ gravity, fitting at the same time the scalaron mass $m$, the mass profile parameters $r_s$ and $r_{200}$ and the anisotropy normalization $\beta_{\infty}$ in the projected radial range $[0.05\,\mpc,r_{200}]$. 
 The 2-dimensional marginalized likelihoods in the plane $(r_{200},\log(m))$,  $(r_s,\log(m))$ and the 1d distribution of the scalaron mass $P[\log(m)]$ are shown in Fig. \ref{fig:marg_1815d} for one halo in the sample; the behavior is equivalent for all the other synthetic clusters.
In each  2-dimensional distribution the red contours indicate points lying within $\Delta\chi^2=2.3$ from the \emph{MAMPOSSt} best fit, while the black solid contours correspond to $\Delta\chi^2=4.6$.
Note that GR is virtually recovered for $\log(m) \gtrsim 1.8$, which corresponds to $\lambda \lesssim 0.015\,\mpc$. This limit has been set considering that for larger $\log(m)$ the relative change in the marginalized likelihood is on average less than $0.1\%$.
The results from \emph{MG-MAMPOSSt} are similar to what found by the dynamical analysis of ref. \cite{Pizzuti17} over the CLASH clusters MACS 1206 and RXJ 2248. This shows that, even in the ideal case, no constraints can be obtained on the scalaron mass from dynamics alone. This fact is a consequence of the degeneracy between the mass profile parameters and the modified gravity parameter in this particular class of models. The velocity dispersion of a self-gravitating system of particles in GR can be mimicked by a modification of gravity with a suitable combination of $r_{200},\,r_s$ and $m$ producing the same effect on the phase space of the tracers. In particular,  from the structure of eq. \eqref{eq:phimod} one can see that an increase of the value of $\lambda=1/m$ (i.e. larger departure from GR) corresponds to an increment in $\phi_{mg}$ which can be compensated by lowering $r_{200}$. In the same way, a shift from GR could be obtained by increasing $r_s$; this is responsible for the peak in the marginalized distribution of $\log (m)$. Moreover, since the term $m\,r_s$ always appear as an exponent, a small variation in the scale radius is sufficient to move the maximum of $P(\log (m))$ away from the GR expectation $m\to\infty$. Nonetheless, GR is always included within $68\%$ C.L., and the results are independent from the values of $r_s$ and $r_{200}$; we remark that the same behavior has been found for all the phase spaces analyzed.\\
We now simulate the possibility to include additional information on $r_s$ and $r_{200}$ as a bivariate Gaussian distribution:
\begin{displaymath}
 P_L(r_s,r_{200})=\frac{1}{2\pi\sigma_{r_s}\sigma_{r_{200}}\sqrt{1-\rho^2}}\exp\left\{-\frac{1}{2(1-\rho^2)}\left[\frac{(r_s-\bar{r}_s)^2}{\sigma^2_{r_s}}+\right.\right.
\end{displaymath}
\begin{equation}
\left.\left.+\frac{(r_{200}-\bar{r}_{200})^2}{\sigma^2_{r_{200}}}-\frac{2\rho(r_s-\bar{r}_s)(r_{200}-\bar{r}_{200})}{ \sigma_{r_s}\sigma_{r_{200}}} \right]\right\},
\end{equation} centered on the true values of the NFW parameters $\bar{r}_s,\bar{r}_{200}$ shown in Table \ref{tab:perfc}. In the above equation, $\rho$ indicates the correlation. Setting those kind of priors on the parameters of the NFW model has the scope of mimicking the information provided by lensing mass profile reconstruction. We then obtain the joint likelihood distribution $\log \mathcal{L}_{tot}=\log \mathcal{L}_{dyn}+ \log P_L$, where $\mathcal{L}_{dyn}(r_{200},r_s,\beta,m)$ is the likelihood from the \emph{MG-MAMPOSSt} analysis. In general, the virial radius $r_{200}$, which is related to the total cluster mass, can be constrained much better than the shape of the halo mass profile, expressed in terms of $r_s$. As shown e.g. in Table 2 of ref. \cite{Umetsu16}, typical uncertainties on the scale radius given by a lensing probe are of the order of $\sim30\div40\%$, while $r_{200}$ can be recovered up to $\sim 5\div10\%$; we thus assume a fixed $\sigma_{r_s}=0.4\times r_s$, while we change $\sigma_{r_{200}}$ to investigate the variation of the bounds on the modified gravity parameter  as a function of the additional constraints on the virial radius. As for the correlation $\rho$ between $r_s$ and $r_{200}$ in $P_L(r_s,r_{200})$ we use the value $\rho=0.67$ found by fitting a bivariate Gaussian on the posterior distribution of the Strong+Weak lensing analysis of MACS 1206 (see again ref. \cite{Pizzuti19} and also refs. \cite{Umetsu16,Caminha2017}). It is worth to notice that, despite different values of the correlation mildly change the shape of the confidence regions for each cluster, the overall qualitative behavior resulting from our analysis is independent of the choice of $\rho$. \\ 
In Fig.  \ref{fig:speriamo} we plot the contours at $\Delta\chi^2=2.3$ in the 2-dimensional plane $(r_{200},\log(m))$ (left) and $(r_s,\log(m))$ (right) for the same halo of Fig. \ref{fig:marg_1815d},  increasing the strength of the prior on $r_{200}$. \\
This simple exercise shows that an additional information on $r_s,\,r_{200}$ with $40\%$ and $7\div10\%$ uncertainties respectively is on average sufficient to produce constraints on $m$ by the joint "lensing"+dynamics analysis of a single ideal cluster when the number of galaxies in the dynamics analysis is $\sim 10^3$. It is worth to notice that this result doesn't change significantly with the number of dynamical tracers in the phase space; moreover, each synthetic phase space never excludes the GR limit within $68\%$ C.L. 
The dependence of the averaged lower limits in the marginalized distributions of $m$ on the strength of the prior on $r_{200}$  at $\Delta \chi^2=1.0$ (red bars) and $\Delta \chi^2=4.0$ (blue bars)  is  shown in the left panel of Figure \ref{fig:lim_lensm}. Each point is obtained by averaging the lower limits for every halo, while the associated errorbars correspond to the dispersion around the mean value.\\
 In order constrain $m$ at $95\%$ confidence level with one cluster, one need at least $\sigma_{r_{200}}=0.07\times r_{200}$. The right panel of the same Figure displays instead the distributions obtained by combining the marginalized total likelihoods of all the 15 clusters as a function $\sigma_{r_{200}}$. In this case, a prior of $40\%$ in $r_s$ and $r_{200}$ for a relatively small number of halos is already enough to provide stringent bounds on the scalaron mass. We obtain: \begin{equation*}
m\ge 7.6\, \mpc^{-1}\,\,\,\,\,\,\,\,\,\Delta\chi^2=4.0,
\end{equation*}
\begin{equation}
m\ge 18.8\, \mpc^{-1}\,\,\,\,\,\,\,\,\,\Delta\chi^2=1.0,
\end{equation}
corresponding to $\lambda\le0.053\,\mpc$ at $\Delta\chi^2=1.0$ which is close to the lower limit of the radial range considered in this analysis ($R=0.05\,\mpc$), and thus it is the tightest constraint reachable with our method. Indeed, decreasing $\sigma_{r_{200}}$ produces negligible effects on the combined distribution, as shown in Fig. \ref{fig:lim_lensm}; clearly, this result relies on the perfect control of the assumptions in the analysis.

\end{document}